\documentclass[fleqn,usenatbib]{mnras}

\usepackage{newtxtext,newtxmath}
\usepackage{units}

\usepackage[T1]{fontenc}

\DeclareRobustCommand{\VAN}[3]{#2}
\let\VANthebibliography\thebibliography
\def\thebibliography{\DeclareRobustCommand{\VAN}[3]{##3}\VANthebibliography}

\usepackage{graphicx}	
\usepackage{amsmath}	
\usepackage{subfig}
\usepackage{units}

\title[Jointly analysing kilonovae and afterglows]{A detailed dive into fitting strategies for GRB afterglows with contamination: A case study with kilonovae}

\author[W. F. Wallace and N. Sarin]{
Wendy F. Wallace,$^{1, 3, 4, 5, 6}$\thanks{wendy.f.wallace@bath.edu}
Nikhil Sarin,$^{2, 3}\thanks{nikhil.sarin@su.se}$
\\
$^{1}$Department of Physics, University of Bath, Claverton Down, Bath, BA2 7AY, UK \\
$^{2}$The Oskar Klein Centre, Department of Physics, Stockholm University, AlbaNova, SE-106 91 Stockholm, Sweden\\
$^{3}$Nordita,
Stockholm University and KTH Royal Institute of Technology
Hannes Alfvéns väg 12, SE-106 91 Stockholm, Sweden\\
$^{4}$Southeastern Universities Research Association (SURA), 1201 New York Avenue NW, Suite 430, Washington, DC 20005 USA\\
$^{5}$Astrophysics Science Division, NASA Goddard Space Flight Center, Greenbelt, MD 20771, USA\\
$^{6}$Center for Research and Exploration in Space Science and Technology, NASA/GSFC, Greenbelt, MD 20771}

\date{Accepted XXX. Received YYY; in original form ZZZ}

\pubyear{2024}

\begin{document}
\label{firstpage}
\pagerange{\pageref{firstpage}--\pageref{lastpage}}
\maketitle

\begin{abstract}
Observations of gamma-ray burst (GRB) afterglows have begun to readily reveal contamination from a  kilonova or a supernova. This contamination presents significant challenges towards traditional methods of inferring the properties of these phenomena from observations. Given current knowledge of kilonova and afterglow modelling, observations \textemdash{}as expected\textemdash{} with near-infrared bands and at early observing times provide the greatest diagnostic power for both observing the presence of a kilonova and inferences on its properties in GRB afterglows. However, contemporaneous observations in radio and X-ray are critical for reducing the afterglow parameter space and for more efficient parameter estimation. We compare different methods for fitting joint kilonova and afterglow observations under different scenarios. We find that ignoring the contribution of one source (even in scenarios where the source is sub-dominant) can lead to significantly biased estimated parameters but could still produce great light curve fits that do not raise suspicion. This bias is also present for analyses that fit data where one source is ``subtracted''. 
In most scenarios, the bias is smaller than the systematic uncertainty inherent to kilonova models but significant for afterglow parameters, particularly in the absence of high-quality radio and X-ray observations. Instead, we show that the most reliable method for inference in any scenario where contamination can not be confidently dismissed is to jointly fit for both an afterglow and kilonova/supernova, and showcase a Bayesian framework to make this joint analysis computationally feasible. 
\end{abstract}



\section{Introduction}
Gamma-ray bursts (GRBs) are broadly classified as either short or long depending on the duration they emit $90\%$ of their energy, T$90$; short for T$90 \lesssim \unit[2]{s}$ and long for longer durations. Following the high-energy gamma-ray emission, the relativistic jet collides with the interstellar medium (ISM), producing a broadband afterglow powered (primarily) by synchrotron radiation~\citep{Sari_Piran_Narayan}. Depending on the type of progenitor system, one may also observe a thermal transient in conjunction with the afterglow, namely a kilonova or supernova. The former are associated with GRBs from compact object merger progenitors and are powered by the radioactive decay of rapid capture (r-process) elements~\citep{Lattimer74,li98}. Meanwhile the latter are associated with core-collapse of massive stars and are powered (largely) by nickel decay~\citep{arnett_sne}. 

The smoking gun multi-messenger gravitational-wave and electromagnetic observations of binary neutron star (BNS) merger, GW170817 and the associated short-duration GRB, GRB170817A, confirmed that binary neutron star mergers are the progenitors of some short GRBs~\citep{ligo_grb170817A,grb170817_1,grb170817_2}. A dedicated follow-up campaign would later reveal the presence of a kilonova, AT2017gfo, confirming BNS mergers as a site of r-process nucleosynthesis~\citep[e.g.,][]{ligo_multimessenger,threeknmodels, emcounterpart_pt1, emcounterpart_pt2,em_counterpart, kn_at2017gfo}. Although GRB170817A seemingly provides support for classifying GRBs by their duration and progenitor, recent peculiar GRBs like GRB211211A~\citep{GRB211211A,Troja_GRB211211A} and GRB230307A~\citep{GRB230307A,Yang_GRB230307A}, which are of long duration but appear to be accompanied by a kilonova, alongside previous examples of failed classification based on duration~\citep[e.g.,][]{Levesque2010, Ahumada2021}, suggest GRB classification is not straightforward.

Further understanding the nature of the GRB progenitor requires us to accurately determine the properties of the associated kilonova/supernova and afterglow from different observations. 
This requires developing accurate models for these phenomena and robust inference techniques. For example, parameter estimation of kilonova ejecta properties can play a critical role towards the study of neutron star (NS) post-merger evolution~\citep[e.g.,][]{Sarin_review}, constraints on the nuclear equation of state~\citep[e.g.,][]{raaijmakers}, and better understanding of nuclear physics in kilonovae~\citep[e.g.,][]{barnes16, Metzger_kilonovae, heating_rate}.

As the recent observations of GRB211211A and GRB230307A indicate, afterglow and kilonova observations can occur coincidentally, where the observed data is a superposition of the two individual transient emission mechanisms. Disentangling the contaminated signal is crucial to ensure accurate estimation of parameters and correct physical interpretations. Therefore, it is necessary to establish robust methods for inferring properties from observations. Such efforts are complementary to studies such as \cite{LSST_kne_estimation} that investigate observing strategies in the run-up to the Legacy Survey of Space and Time (LSST)~\citep{LSST}.

In this paper, we highlight the pitfalls of different strategies for parameter estimation of potentially contaminated afterglow and kilonova signals and showcase a more robust framework for parameter estimation. 
In particular, we compare the most common analysis technique where the afterglow and/or kilonova is fit separately or on data where the result of the fit is ``subtracted''~\citep[e.g.][]{GRB050709,GRB160821B,GRB211211A}, to the use of a combined model to fit the kilonova and afterglow simultaneously~\citep[e.g.][]{GRB230307A}. While we focus here on kilonova and GRB afterglows, the general techniques described here are directly applicable to other scenarios where two (or more) emission mechanisms appear together.

This paper is structured as follows, in Sec.~\ref{sec:methods}, we describe the emission models for the afterglow and kilonova and our model combining the two emission mechanisms. In Sec.~\ref{sec: Regime of Contamination}, we show light curves for a population of afterglows and kilonovae, discuss the regime where observations are most likely to be contaminated, and identify wavelengths and epochs with high diagnostic power. We then go on to compare the methods that can be used to disentangle the transients and infer their properties in Sec.~\ref{sec:simulation}, further highlighting the importance of different observing bands and the impact they have on distinguishing between kilonovae and afterglows using photometry alone, and conclude in Sec.~\ref{sec: conclusions}.

\section{Models}\label{sec:methods}
GRB afterglows are well described by synchrotron emission produced from the interaction of a relativistic jet with the surrounding interstellar medium. Observations of GRB170817A confirmed the notion that such jets are structured, which could include an inhomogeneous distribution of energy within the jet cone or emission wings that extend beyond the ultra-relativistic core. However, the exact details of jet-structure are uncertain~\citep{Granot_offax,threeagmodels}. For simplicity, we only consider ``tophat'' afterglow jets in this work following~\citet{tophat_redbackmodel}, which assumes the jet is a narrow ultra-relativistic core defined by an opening angle $\theta_{\rm c}$, outside of which the energy drops to zero. We note that the choice of a ``tophat'' model is likely to well describe a bulk of the parameter space where contamination from a kilonova is most likely to occur: GRBs viewed on-axis. 

Details of kilonova modelling are far less certain. We consider two commonly used models in the literature; a multi-component kilonova model following~\citet{threeknmodels}, which we refer to as a ``two\_component\_kilonova'' model and the ``two\_layer\_stratified\_kilonova'' model following \citet{heating_rate}. The bulk of the ingredients in these commonly used models are largely similar, such as the input heating rates~\citep{korobkin12}. However, the models differ most significantly in their treatment of thermalization, with the former following~\citet{barnes16}, and the latter using an independent treatment following~\citet{heating_rate}. These models also differ in their treatment of the dynamical evolution of the ejecta, with the former following an ``arnett'' model for supernovae~\citep{arnett_sne}, with appropriate modifications to account for r-process nucleosynthesis~\citep{korobkin12, barnes16} and two independent components. The latter model assumes the total mass to be distributed between shells undergoing homologous expansion. 

\begin{figure}
    \centering
    \includegraphics[width=\linewidth]{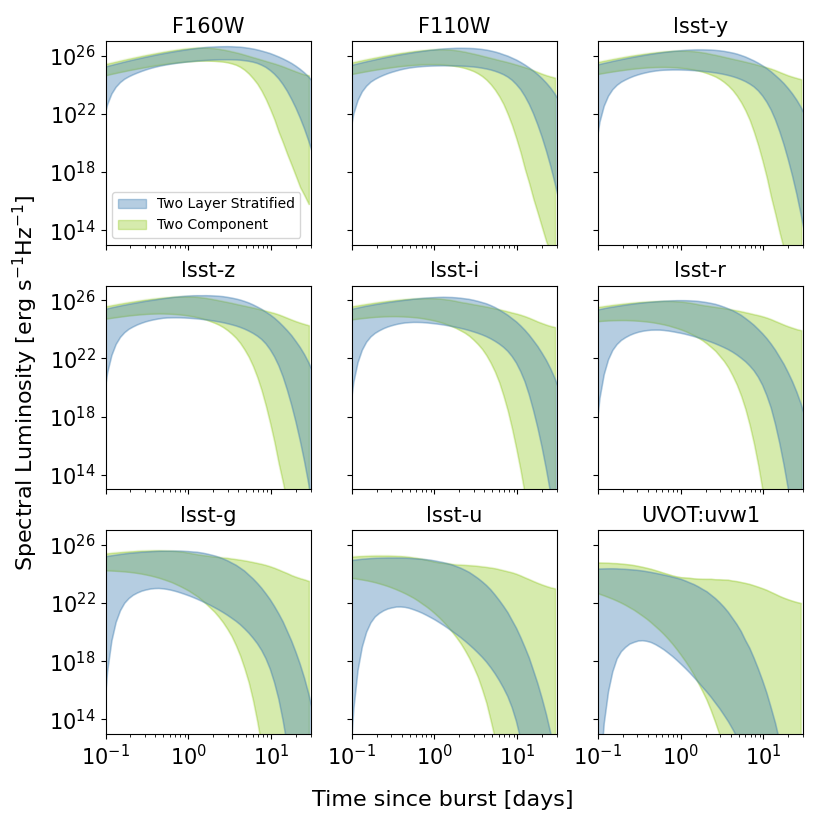}
    \caption{90\% credible interval light curves of two different kilonova models as viewed in different bandpasses.}
    \label{fig:kn_model_comparison}
\end{figure}

In Fig.~\ref{fig:kn_model_comparison}, we plot the 90\% credible intervals for random draws made from each of the kilonova model priors, listed in Tabs.~\ref{tab:kn_twocomp_priors} and~\ref{tab:kn_priors}, which are motivated by inferences of the kilonova AT2017gfo for various observational bands. The models show the largest discrepancies when viewed at bluer wavelengths and at later times. The two-component model generally shows greater variation and later decay times than the ``two\_layer\_stratified'' model. We also note that these models do not capture viewing-angle dependence. However, the effect of viewing angle on kilonova light curves is likely significantly smaller than the viewing angle effect on afterglows. 

Kilonova light curves are far from certain, and inferences of parameters are extremely sensitive to the model choice and the inherent systematics~\citep{sarinrosswog24}. However, given recent works highlight the important role of the dynamical evolution of the ejecta in shaping the lightcurve~\citep{velocity_magnetar, velocity_distrib}, we focus on the ``two\_layer\_stratified\_kilonova'' model for the remainder of this work as it considers the dynamics in a more self-consistent manner. 

The recent observations of GRB211211A and GRB230307A demonstrate that both kilonovae and afterglows can be observed at the same time and create a marked effect on the overall light curve. Meanwhile, numerical simulations have also shown that interaction of the jet with the kilonova ejecta can change the properties of the ejecta, creating significant observable differences to the light curve~\citep{Klion2021, Nativi2021}. For our analysis, we assume that the kilonova and afterglow emission are independent and the observed emission is a linear combination. This simplistic approach is likely unsuitable for on-axis observers, where jet-ejecta interactions likely make the ejecta more energetic and, therefore, `bluer'. However, this is not expected to change our conclusions about the ideal methodology for inferring properties. We discuss limitations of our modelling in greater detail in Sec.~\ref{sec: conclusions}. 

\section{Contamination regime}
\label{sec: Regime of Contamination}
The specific time of observations, bandpasses, and model parameters all ultimately impact the afterglow or kilonova lightcurve. We first investigate the parameter space where a kilonova is likely to contaminate observations of GRB afterglows. We treat on and off-axis afterglow light curves separately in this section due to their distinct appearances. Moreover, the former is more likely to be associated with a prompt GRB and accompanied by X-ray and radio observations, alleviating some of the modelling concerns as we discuss later. 

As the distinct behaviour of on-axis and off-axis observers suggests, the degree of contamination from the kilonova to the afterglow is ultimately sensitive to the afterglow peak timescale. This timescale, in turn is sensitive to the core opening angle $\theta_c$, jet energy $E_0$, and interstellar medium density $n_0$~\citep{Granot_offax}. Similarly, the kilonova peak timescale (dependent on the opacity, mass, and velocity) also plays a role. However, kilonova light curves have more significant variation between observational bandpasses, such that the dependence on the peak timescale itself is minimal compared to the afterglow. 

\begin{figure*}
  \centering
  \subfloat{\includegraphics[width=0.5\textwidth]{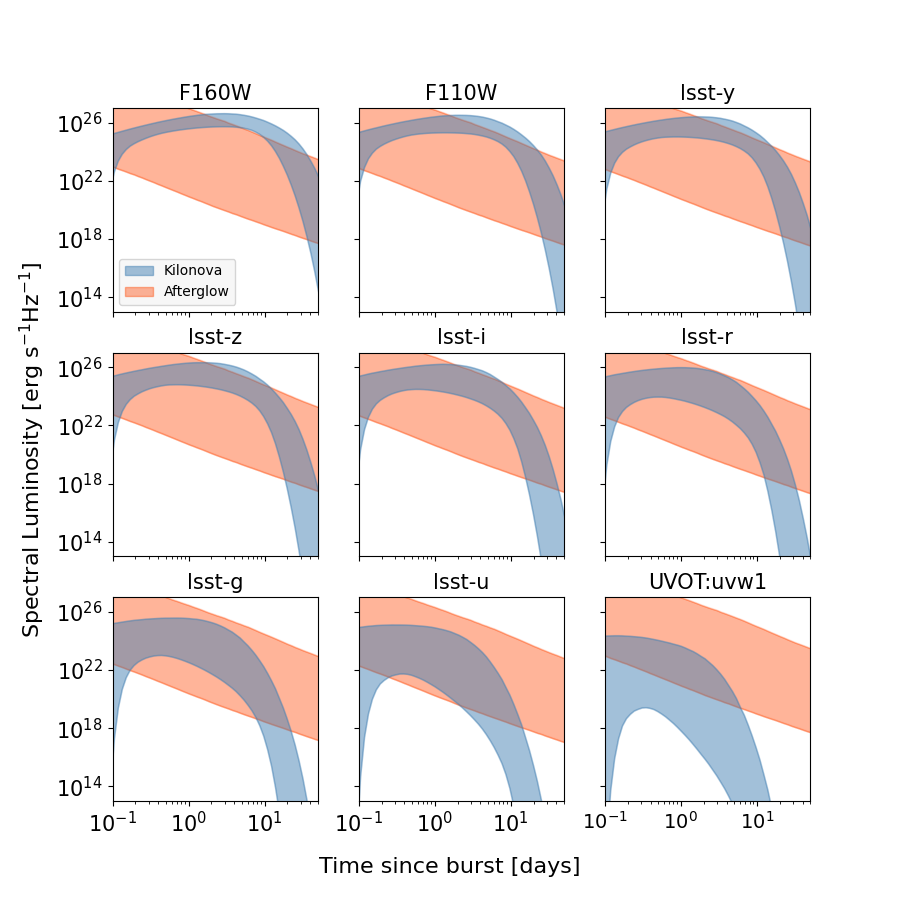}}
  \subfloat{\includegraphics[width=0.5\textwidth]{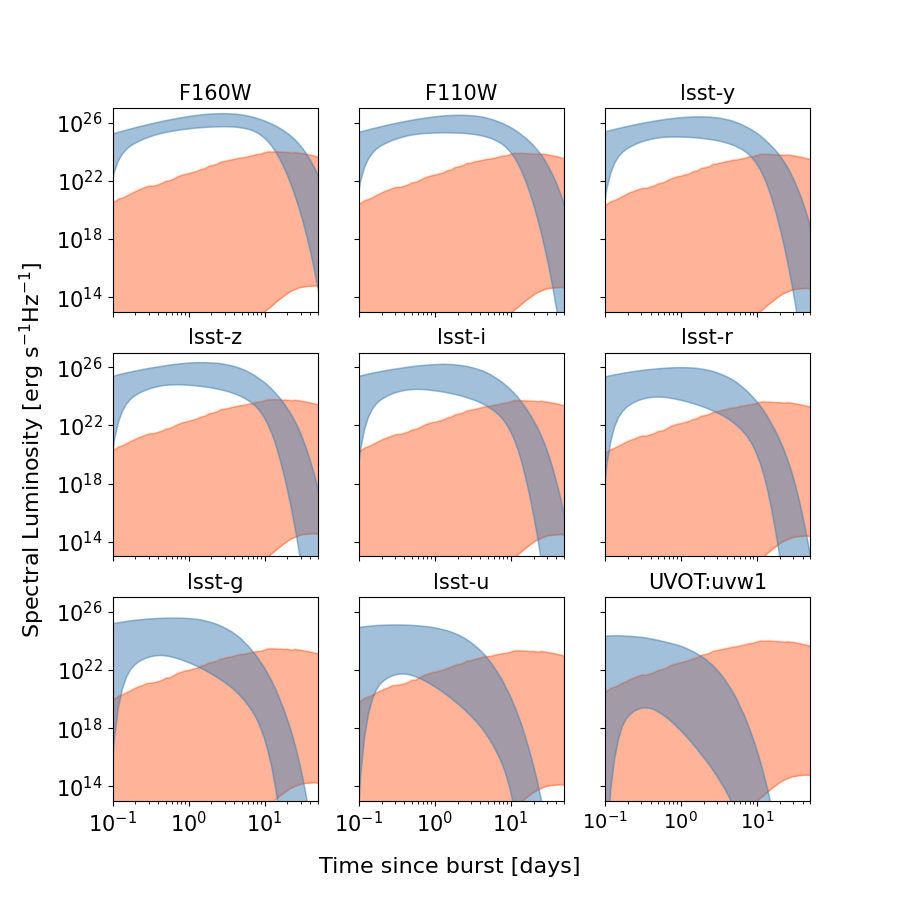}}
  \caption{90\% credible interval light curves in different bandpasses for a kilonova +  afterglow system viewed on-axis (left) and off-axis (right).} \label{fig:model_contamination}
\end{figure*}

\begin{figure*}
  \centering
  \subfloat{\includegraphics[width=0.45\textwidth]{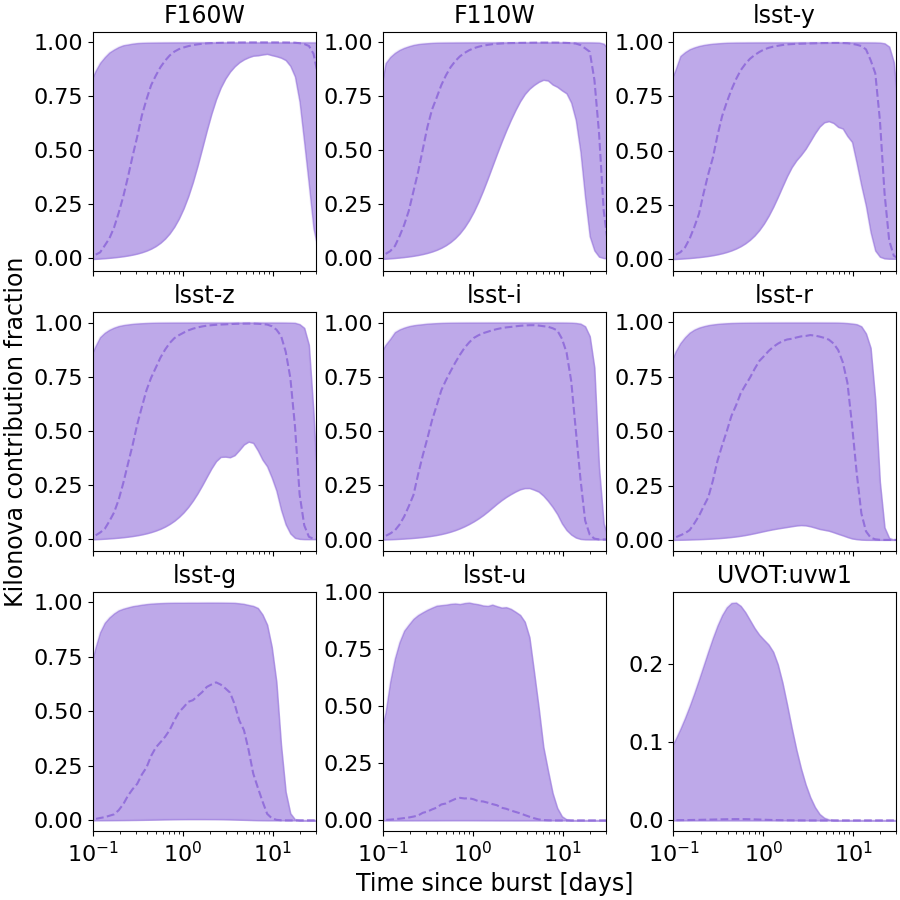}}
  \hspace{0.04\textwidth}
  \subfloat{\includegraphics[width=0.45\textwidth]{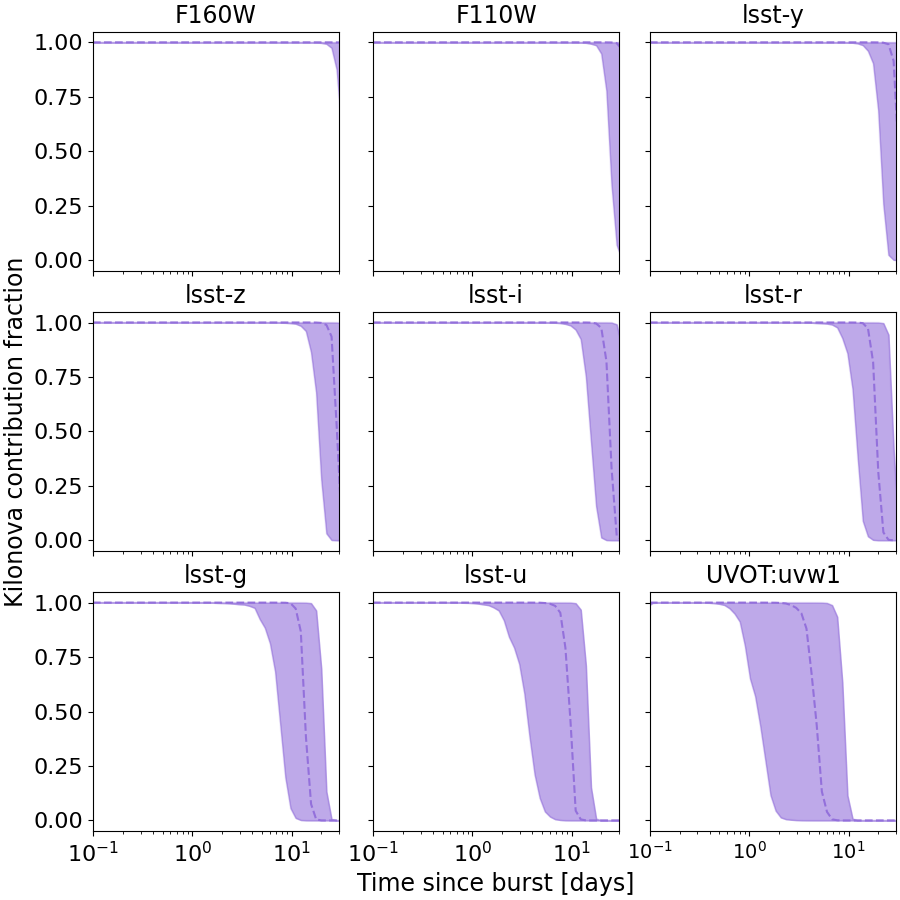}}
  \caption{68\% credible intervals showing the fractional contribution of kilonova flux evolution, relative to total flux observed in different bandpasses, for a kilonova + afterglow system viewed on-axis (left) and off-axis (right). The dashed line indicates the median.} \label{fig:contamination_frac}
\end{figure*}

To demonstrate the possible degree of contamination, we randomly select 5000 kilonova and afterglow parameter sets from priors motivated by inferences of the kilonova, AT2017gfo~\citep{threeknmodels}, and afterglow of GRB170817A~\citep{threeagmodels} all at the same distance of 44.6 Mpc. We select parameters twice for the afterglow, once to represent an on-axis observation scenario when the viewing angle is within the jet cone; $\theta_v < \theta_c$, and once to represent an off-axis scenario when viewing outside the jet; $\theta_v > \theta_c$. The 90\% credible intervals for resulting light curves are shown in Fig.~\ref{fig:model_contamination}, and we specify the parameters and priors of the kilonova and afterglow models in Tabs.~\ref{tab:kn_priors} and~\ref{tab:ag_priors}, respectively. 
In Fig.~\ref{fig:contamination_frac}, we explicitly show how the fractional contribution of flux from the kilonova, relative to the combined flux as expected from Fig.~\ref{fig:model_contamination}, evolves with time, and show the median and 68\% credible intervals in each subplot. In addition to near-infrared (NIR) and ultraviolet (UV) bands, we include all LSST filters to highlight the degree of contamination to the afterglow at different wavelengths for an on-axis (off-axis) afterglow in the left (right) panel. From the figures, we immediately see strong dependence on the viewing angle $\theta_v$, with on-axis systems likely suffering from significant contamination at all wavelengths. Meanwhile, systems viewed off-axis are only susceptible to significant contamination at late-times ($t \gtrsim \unit[10]{d}$). As a specific example, we can consider the case of GRB170817, which was observed at least $\gtrsim \unit[20]{\deg}$ off-axis, and there the kilonova had no discernible impact on the afterglow.

\begin{table}
    \centering
    \begin{tabular}{ccc}
         \hline
         Parameters&  Distribution& Range\\
         \hline
         Redshift, $z$& Fixed & (0.01)\\  
         Ejecta mass 1(2), $M_{\rm ej-1(2)}$&  Uniform& (0.01, 0.03) $M_\odot$\\
         Velocity 1(2), $v_{\rm ej-1(2)}$&  Uniform& (0.1, 0.5) $c$\\
         Grey opactity 1(2), $\kappa_{1(2)}$&  Uniform& (1, 30) cm$^2$/g\\
         Temperature floor 1(2), $T_{\rm floor-1(2)}$& LogUniform& (100, 6000) $K$\\
    \end{tabular}
    \caption{Priors for component 1(2) in the ``two\_component\_kilonova\_model'' used for random draws, where $M_\odot$ is units of solar mass and $c$ is the speed of light.}
    \label{tab:kn_twocomp_priors}
\end{table}

\begin{table}
    \centering
    \begin{tabular}{ccc}
         \hline
         Parameters&  Distribution& Range\\
         \hline
          Redshift, $z$& Fixed& {(0.01)}\\  
         Ejecta mass, $M_{\rm ej}$&  Uniform& (0.01, 0.05) $M_\odot$\\
         Inner shell velocity, $v_{\rm ej-1}$&  Uniform& (0.05, 0.2) $c$\\
         Outer shell velocity,  $v_{\rm ej-2}$&  Uniform& (0.3, 0.5) $c$\\
         Grey opactity, $\kappa$&  Uniform& (1, 30) cm$^2$/g\\
 Density profile power law index, $\beta$& Uniform&(3.1, 8)\\
    \end{tabular}
    \caption{``two\_layer\_stratified\_kilonova'' priors used for random draws and parameter estimation.}
    \label{tab:kn_priors}
\end{table}

\begin{table}
    \centering
    \begin{tabular}{ccc} 
    \hline
     Parameter & Distribution & Range \\
     \hline
     Redshift, $z$& Fixed& 0.01\\  
     Viewing angle, $\theta_v$& Sine& (0, $\pi$/2) rad\\
     Jet half opening angle, $\theta_c$& Uniform&(0, 0.1) rad\\
     Isotropic jet energy, $\log_{\rm 10}(E_0)$& Uniform&(44, 54)\\
     ISM density, $\log_{\rm 10}(n_0)$& Uniform&(-5, 2) \\
     Accelerated electron fraction, $\chi_N$& Fixed&1\\
     Electron spectral index, $p$& Fixed&2.3\\
     Fraction of energy in electrons, $\epsilon_e$& Fixed&-1.25\\
     Fraction of energy in magnetic field, $\epsilon_b$& Fixed&-2.5\\
     Initial Lorentz factor, $\Gamma_0$& Fixed&1000\\
  \end{tabular}
  \caption{``tophat\_redback'' afterglow priors used for random draws}
  \label{tab:ag_priors}
\end{table}

Investigating the specific on-axis afterglow case further, we generally find that the possibility of significant kilonova contamination strongly depends on the overall brightness of the afterglow. In particular, we find significant contamination to the afterglow when it is brighter than $\unit[2.4\times10^{\rm 20}]{erg~s^{-1}~Hz^{-1}}$ at $\unit[1]{day}$ post GRB, in the lsst-r band. Inspecting the left panels of Figs.~\ref{fig:model_contamination} and \ref{fig:contamination_frac}, we see that bandpasses at higher effective wavelengths are more likely to have significant contamination, a direct consequence of the competition between the non-thermal GRB and thermal (but rapidly cooling) kilonova spectrum. In particular, almost all NIR to optical wavelengths have significant periods where the kilonova spectral luminosity contributes almost entirely to the total spectral luminosity across all of our $5000$ random draws. However, at bluer wavelengths e.g., UVOT:uvw1 filter, the median kilonova contribution is negligible at all times, never exceeding $\gtrsim$ 0.3\%, suggesting that this filter offers the best opportunity (at near optical wavelengths) to probe just the afterglow. The 90\% credible intervals in Fig.~\ref{fig:model_contamination} show that contamination is likely in all wavelengths, however at longer wavelengths where the kilonova evolves the slowest, it is possible to identify the transient as a bump superimposed on the afterglow lightcurve. Together, Figs.~\ref{fig:model_contamination} and~\ref{fig:contamination_frac} suggest that wavelengths from NIR to lsst-r should hold the greatest diagnostic power for kilonova detection and subsequent parameter estimation, with the optimal observation window ranging between approximately $\unit[2-15]{days}$. However, we do emphasize that these statements are based on the broad interpretation of our randomly sampled population, and specific cases may be significantly different. 

Much more so than the on-axis scenario, contamination between an off-axis afterglow and a kilonova is strongly sensitive to the afterglow peak timescale and brightness, with earlier and brighter peaks more likely to contribute to signal overlap. For our choice of distance, we find significant contamination to only occur where the peak spectral luminosity is greater than $\unit[2.4\times10^{\rm 15}]{erg~s^{-1}~Hz^{-1}}$ and occurs at a time before $\unit[200]{days}$ post GRB in the lsst-r band. 
The impact of observational bandpasses on kilonova contamination to an off-axis GRB afterglow is shown in the right panels of Figs.~\ref{fig:model_contamination} and~\ref{fig:contamination_frac}. We draw attention to the certainty of kilonova domination at early times. 
In the NIR bands, we see that the kilonova contributes almost all of the observable flux up to at least $\unit[30]{days}$ post GRB, but this window can decrease to as little as $\unit[1]{day}$ in the UV. After such times, we see the kilonova contribution fall rapidly due to both the onset of kilonova decay and the afterglow rising to peak, as depicted in Fig.~\ref{fig:model_contamination}. This further emphasises the high diagnostic power of NIR bands and the importance of early observation times for kilonova detection and parameter estimation. Although this bodes well for the kilonova, it may hinder the afterglow recovery, especially as in an off-axis scenario we are less likely to have X-ray or radio coverage, which is more likely triggered by the detection of a GRB. As we show in more detail in Sec.~\ref{sec: scenario 3}, the lack of radio and X-ray data weakens our ability to accurately infer the parameters of the kilonova and afterglow.  

\section{Simulations}
\label{sec:simulation}
We now explore different strategies for fitting observations under different contamination/observational scenarios. In particular, we apply different fitting techniques to simulated data sets, ensuring we know all aspects of the underlying model, noise, and true input parameters. The data represents contamination scenarios involving both on- and off-axis afterglows and situations with and without X-ray and radio observations. The afterglow and kilonova are fit individually, using the same transient model as the input, and jointly, using a combined transient model. Individual kilonova fits are also made to data that we first subtract the independently inferred afterglow from. By comparing the estimated values and fits of these different scenarios with the input values used to simulate the relevant data set, we aim to understand what the ideal method and observational practices are for robust parameter estimation and interpretation. 

The dominant cost of fitting for all our scenarios was the cost of the afterglow model, where each evaluation of the model could take up to $\unit[500]{ms}$. To reduce this cost, we first develop a machine-learning surrogate model for the afterglow model using a multi-layer perceptron (MLP) neural network, with the {\sc scikit-learn} package. The model hyperparameters are chosen such that the network is composed of three hidden layers, each containing 300 perceptrons, with the rectified linear unit function as the non-linear activation function and adam as the optimizer~\citep{adam}. We use 743,985 parameter samples and corresponding flux density values generated with our actual afterglow model, with a train-test split of 0.15/0.85. Samples are drawn from the unconstrained prior distributions of the tophat afterglow with only redshift, $z=0.01$ and electron participation fraction, $\chi_N=1$ as fixed parameters. To ensure a representative training set, we distinguish between viewer alignments (on-axis and off-axis afterglows) and split the frequency range into nine logarithmic bins. Although redshift is fixed in our samples, we account for distance variations by applying a scaling correction to the flux density within {\sc redback} for other specified values. We pre-process training samples in a similar manner to~\cite{Sarin2021_xt1, deepglow}. This is done by first taking the base 10 logarithm of every parameter, except $p$, and applying the standard scaler function from {\sc scikit-learn} \citep{scikit-learn}. Similarly, we pre-process the target flux density values by taking their natural logarithm and applying the standard scaler function. Note that while the emulator is trained to output flux densities, we have converted to spectral luminosity for the figures in this paper to showcase results in a distance independent format.

After training, we find the test set returns a score of 0.99, indicating the emulator reproduces outputs which are consistent with those generated by the ``tophat\_redback'' model from the same set of parameters. The uncertainties in our surrogate model are within those inherent to afterglow modelling. We further validate the surrogate model, by performing full simulations where we generate data with the true numerical model and perform inference with the surrogate recovering consistent posteriors as with the original model. We find that our emulator is consistently at least 1-2 orders of magnitude faster than the original model depending on the size of the output frequency and time arrays. As our primary aim of this work is to test the methodology for fitting contaminated observations of transients, we ignore the small error in the machine-learning model itself.
We make this model publicly available through the {\sc{Redback}} package.

We explore two primary synthetic data in Secs.~\ref{sec: scenario 1}, \ref{sec: scenario 2} and~\ref{sec: scenario 3}, both representing scenarios where there is significant contamination between a kilonova and an afterglow. Two additional limiting case scenarios that we explore are where the kilonova is sub-dominant (Sec.~\ref{sec: scenario 4}) and where there is an afterglow observation but no true kilonova contamination (Sec.~\ref{sec: scenario 5}). The first scenario in Sec.~\ref{sec: scenario 1} reflects the observations of a system viewed on-axis, and the second in Sec.~\ref{sec: scenario 2} reflects those of a system viewed off-axis. In each case,  unless otherwise specified, we generate 100 synthetic data points at a redshift of $z=0.01$ with normally distributed noise using a sigma that is calculated at 0.25 $\times$ the model output. We apply extinction using a dust extinction function following \cite{fitzpatrick99}, with extinction in the V-band $av=0.5$ and the ratio of total to selective extinction $R_v=3.1$. Although we simulate with the same level of extinction in these scenarios, we keep $av$ as a free parameter during inference ($R_v$ fixed to 3.1) and use a uniform prior with a range of 0 - 2 magnitudes. The observational data is randomly generated for the bands presented in the previous sections as well as at frequencies of $\unit[5\times10^9]{Hz}$ and  $\unit[2\times10^{\rm 17}]{Hz}$, to represent radio and X-ray emission. Between the on-axis and off-axis significant contamination scenarios, we only change three parameters: $\log_{\rm 10}(E_0)= 51.0$, $\log_{\rm 10}(n_0)=0.5$, $\theta_v=0.03$~rad for the on-axis system and $\log_{\rm 10}(E_0)= 51.5$, $\log_{\rm 10}(n_0)=1$, $\theta_v=0.5$~rad for the off-axis system. We specify the parameters for scenarios 4 and 5 in their relevant sections (\ref{sec: scenario 4} and \ref{sec: scenario 5}) . The remainder of the afterglow injection parameters, bar $\theta_v$ where specified, are the same across all scenarios with values of $p=2.3$, $\log_{\rm 10}(\epsilon_e)=-1.25$, $\log_{\rm 10}(\epsilon_b)=-2.5$, $\chi=1$, $\gamma_0=1000$, $\theta_c= 0.07$~rad. Where present, we also keep the kilonova the same with injection values of $M_{\rm ej}=0.03$$~M_{\odot}$, $v_{\rm ej-1}=0.1$$~c$, $v_{\rm ej-2} =0.4$$~c$, $\kappa=5$$~\mathrm{cm}^2/\mathrm{g}$ and $\beta=4$.  

For all sets of data, we compare different strategies to fit the observations and infer the properties of the kilonova and afterglow. 1) We explore fitting the afterglow and kilonova models independently, fitting only one transient at a time to the full set of contaminated data. 2) Joint fitting, where we simultaneously fit the data for both transients with the combined model and 3) A strategy commonly used where we first subtract the best-fitting afterglow, according to an independent afterglow-only fit, from the full data, and then fit the kilonova to the subtracted data.
We fit the observations using the same models as the input with the strategies outlined above and quantify the robustness of these methods more generally in Sec.~\ref{sec: biases}. The fits are performed through {\sc Redback}~\citep{redback_paper} using either the {\sc emcee}~\citep{emcee} or {\sc Nestle}~\citep{nestle} samplers wrapped within {\sc Bilby}~\citep{bilby1}. We primarily use {\sc emcee} to further reduce computational cost, starting our inference with walkers centered on regions of maximum likelihood. However, in cases where MCMC sampling fails to capture multi-modalities we use {\sc Nestle} as nested samplers are better equipped to sample such posteriors.

In Sec.~\ref{sec: scenario 3}, we investigate how the availability of observational data impacts the ability to disentangle the transients. Since the kilonova does not emit at X-ray and radio frequencies, observations of the afterglow in these bands are particularly interesting as they can constrain the afterglow parameter space during inference to make the overall fitting process more efficient. To highlight their importance, individual afterglow and joint fits are made to the same two data sets as described above, but with the X-ray and radio observations removed. It is in such `difficult' scenarios, that using a nested sampler e.g., {\sc Nestle} becomes more necessary, compared to a simpler MCMC sampler.

\begin{figure*}
    \centering
    \includegraphics[width=0.9\textwidth]{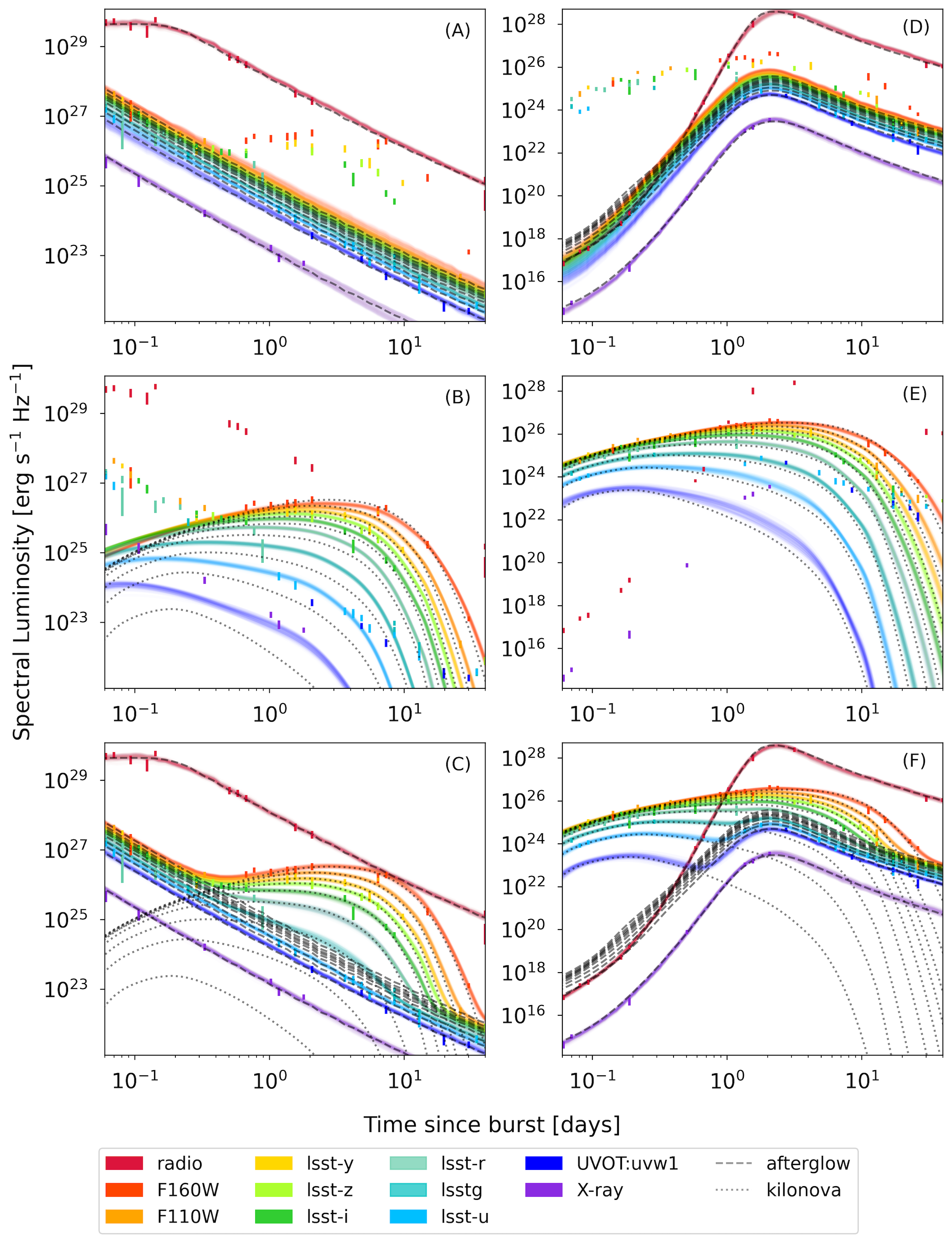}
    \caption{Each panel contains the full set of synthetic data points from either the on-axis contamination case of~\ref{sec: scenario 1} (left; A,B,C) or the off-axis case of~\ref{sec: scenario 2} (right; D,E,F). The figure shows fits made to the full data using the afterglow only model (upper; A,D), the kilonova only model (middle; B,E) and the joint model (lower; C,F). In each panel, 100 solid coloured light curves show posteriors for each fit, where each colour represents a different observation band. The true value curves for the afterglow and kilonova components are overplotted in grey dashed and dotted lines respectively.}
    \label{fig:lightcurves}
\end{figure*}

\subsection{Scenario 1 - On-axis afterglow with significant kilonova contamination}
\label{sec: scenario 1}

As discussed in Sec.~\ref{sec: Regime of Contamination}, an on-axis GRB system is likely to suffer from significant kilonova contamination. As the cases of GRB211211A and GRB230307A  already prove, they are also the most observationally relevant. Our fits using strategies 1 and 2 for this on-axis scenario and the following off-axis scenario are presented in Fig.~\ref{fig:lightcurves}, where each subplot shows 100 randomly drawn light curves from the posterior distributions as solid lines, with colours corresponding to the observation band. For comparison, we overplot the expected curves of the afterglow and kilonova components, based on their injection values, as dashed and dotted lines respectively. 

We first focus on the strategy where the kilonova is fit independently to the data. The fitted light curve from this strategy is shown in Fig.~\ref{fig:lightcurves} (panel B), which highlights the poor recovery relative to the true input. However, we emphasize that in real observation scenarios, true parameter values are not known, so one could not directly assess the accuracy of inference in this manner. If we disregard the true input curves, the posteriors in Fig.~\ref{fig:lightcurves} appear to reasonably match the relevant data points for the transient being fit, suggesting a `good' fit to the data that may lead to the conclusion that the inferred parameters are correct. 

\begin{figure}
    \centering
    \includegraphics[width=\linewidth]{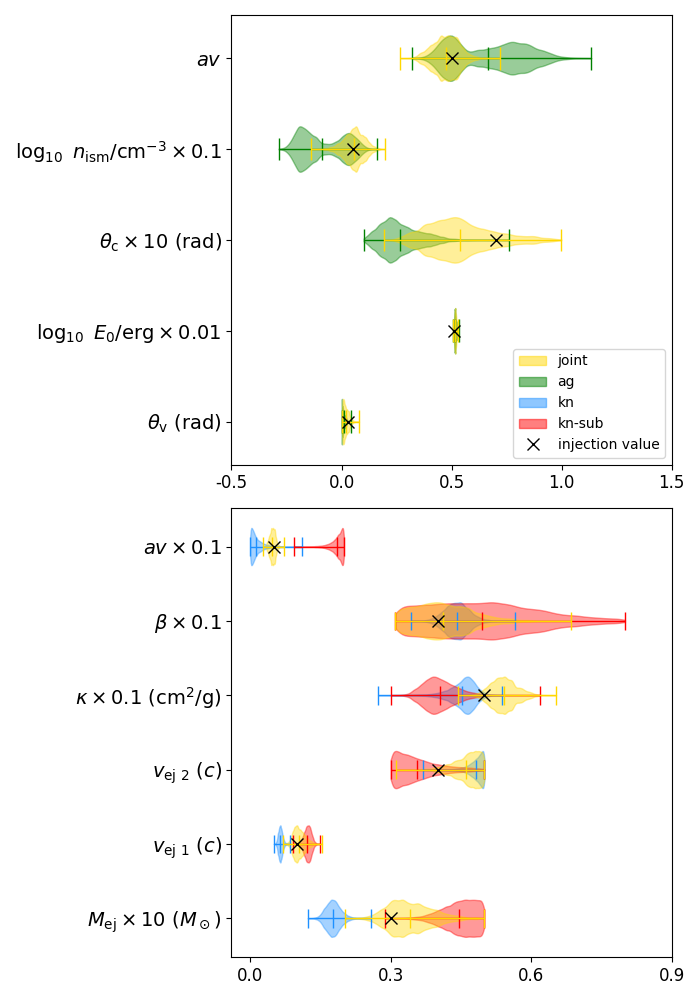}
    \caption{Posterior distributions obtained using the joint (joint), afterglow only (ag), kilonova only (kn) and kilonova after afterglow subtraction (kn-sub) fitting methods for a system viewed on-axis. Parameters for the afterglow are shown in the top panel and those for the kilonova are shown in the bottom panel.}
    \label{fig:violin_onax}
\end{figure}

In Fig.~\ref{fig:violin_onax}, and similar figures within this section, we show violin plots for relevant afterglow parameters in the upper panel and kilonova parameters in the lower panel with individual afterglow fits in green, individual kilonova fits in blue, kilonova fits after afterglow subtraction in red, and joint fits in yellow. Injection values are marked by `$\times$'. Despite producing good visual fits to the data, our inferred parameters from the individual kilonova fit are significantly biased. In particular, $v_{\rm ej-1}$, $M_{ej}$ are biased by $\gtrsim 10\sigma$ and $\gtrsim 6\sigma$, respectively (relative to the statistical errors), and show no posterior overlap with the true input. Given that these parameters are the most important for interpreting a kilonova, the error seen here suggests that solely fitting the kilonova is not a good strategy. 

As shown in Fig.~\ref{fig:lightcurves} (panel A), the independent fit to the afterglow is a more viable strategy for producing a better match to the true input. This is largely attributed to including X-ray and radio data, which provide some ``calibration'' for the afterglow model as the kilonova model has no flux at these wavelengths.  
This motivates the use of the commonly-used strategy of attempting to isolate the kilonova signal by first subtracting the predicted afterglow, and then performing a fit.
We approach this ``subtraction'' strategy by masking the X-ray and radio data points and then subtracting the afterglow contribution from all other wavelengths. We generate the afterglow contribution using our ground truth ``tophat$\_$redback'' afterglow model with the best-fitting model parameters as inputs and no additional error term.  Ideally, the errors on the subtracted data points should account for the errors on both the model and the original data. The former can be estimated using samples from the posterior, and in our study for most of our simulated data, this produces errors significantly smaller than the original simulated data error. We therefore work with the original error to simplify our analyses when accounting for the uncertainty in the subtraction.

While most of the posteriors using the subtraction method, shown in Fig.~\ref{fig:violin_onax}, are more consistent with the input than the independent kilonova-only fit, they are still significantly biased. Notably, again, neither the kilonova ejecta mass nor shell velocity are consistent with the input, highlighting the pitfalls of this commonly used strategy. Additionally, extinction is recovered worse and now shows no posterior overlap with the true input. We note that increasing the size of the (largely arbitrary) uncertainty on the error bars when fitting the afterglow-subtracted data could make the kilonova parameters more consistent with the input. However, from a Bayesian perspective, the ``subtraction'' approach is incorrect and an example of a misspecified model. 
In particular, when we first fit the afterglow model to data, to maximise the likelihood, the afterglow will depart from the `truth', especially, but not only, when radio and X-ray data errors are high or the quality otherwise due to cadence or spectral coverage is low relative to the optical. 
This biased afterglow recovery is then built into the kilonova parameter estimation. 
Since the intrinsic kilonova and afterglow models are themselves uncertain (unlike our simulation where we know the input), the subtraction strategy can cause significant issues with interpretations. By contrast, the proper (joint inference) Bayesian approach to this problem produces kilonova and afterglow posteriors that are marginalising over the other (afterglow) model as opposed to being conditioned on it, and so they will naturally be consistent with the input (given adequate sampling). Moreover, by marginalising over the other model parameters, it is far more likely to yield robust estimated kilonova parameters, as opposed to the subtraction approach. 

To verify that the joint approach works, we show fits to the data in Fig.~\ref{fig:lightcurves} (panel C) and yellow violins for comparison to the other methods in Fig.~\ref{fig:violin_onax}, highlighting consistency with the inputs. A full corner plot for the joint recovery can be found in Fig.~\ref{fig:sig_on_joint_corner} of the appendix, which confirms an accurate fit with all true kilonova and afterglow parameters recovered correctly to within 3$\sigma$ of the predicted values. We also emphasize that the joint-fitting approach is not significantly more computationally intensive than independently performing an afterglow-only and, subsequently, kilonova-only fit. Importantly, it does not require extra data processing to ``generate'' afterglow-subtracted kilonova data. 

\subsection{Scenario 2 - Off-axis afterglow with significant kilonova contamination}
\label{sec: scenario 2}
Following the same analysis approaches as Sec.~\ref{sec: scenario 1}, we again compare individual, afterglow subtracted, and joint fits for an off-axis GRB system. 
In Sec.~\ref{sec: Regime of Contamination} we noted that at early observation epochs, the off-axis afterglow spectral luminosity is typically low, which allows the kilonova to dominate the signal in all wavelengths. This suggests that strategies which involve individually fitting the kilonova to the data may be more suited towards the off-axis contamination scenario than the on-axis scenario. The result of our individually fit kilonova component relative to the true inputs, shown in Fig.~\ref{fig:lightcurves} (panel E), seems to reflect this. 

\begin{figure}
    \centering
    \includegraphics[width=\linewidth]{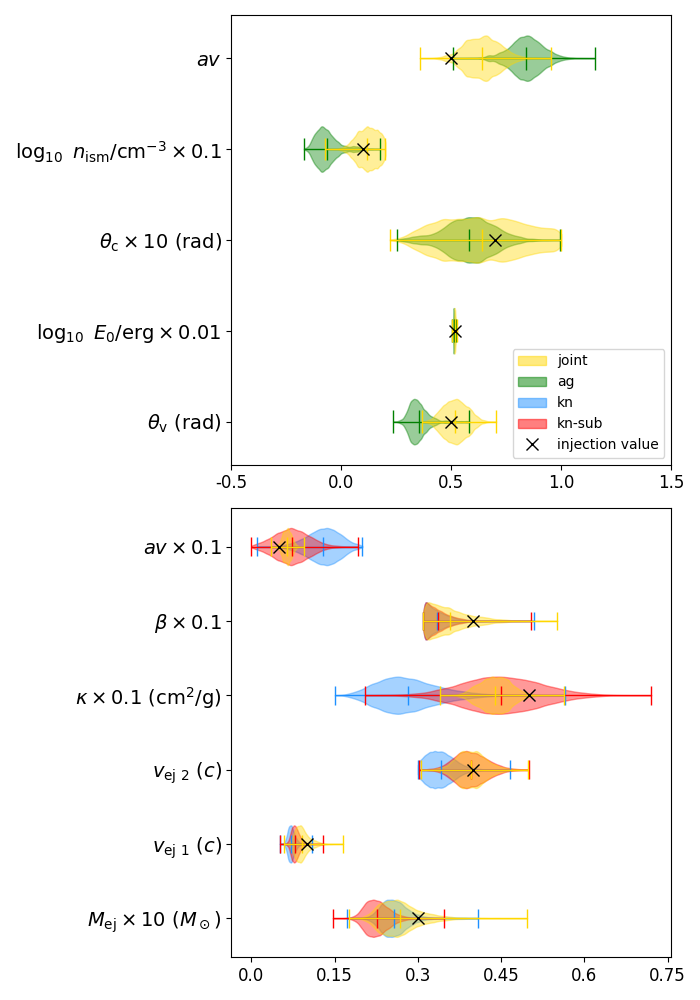}
    \caption{Similar to Fig.~\ref{fig:violin_onax} but for the off-axis scenario.}
    \label{fig:violin_offax}
\end{figure}

In Fig.~\ref{fig:violin_offax}, we see the biases on ejecta mass and shell velocity are reduced in comparison to the on-axis GRB case, discussed in Sec.~\ref{sec: scenario 1}. 
However, the posterior of $v_{\rm \rm ej-1}$ remains inconsistent with our input. We additionally see that opacity is incorrectly inferred, which is likely due to sensitivity to contamination around the knee of the lightcurve, as this is typically the source of constraints on opacity (alongside the peak time) for the kilonova models used in this work. The inaccuracy of the recovered posteriors, despite a seemingly good fit to relevant data points, highlights the need for caution in interpreting fits to subtracted data, and general caution for detailed parameter estimates of quantities with this fitting strategy. 

We now look at the individual afterglow fit to data and note that while the kilonova is largely dominant in the lower frequency bandpasses, up to 10 days in lsst-r, this is not a complete detriment to the afterglow recovery. This is again due to the inclusion of uncontaminated radio and X-ray observations which provide a level of calibration to the fit. In the upper panel of Fig.~\ref{fig:violin_offax}, we show green violin plots for the key afterglow parameters which we identified as significantly influential to contamination in Sec.~\ref{sec: Regime of Contamination}, and additionally extinction. Most of these properties are recovered at a 3$\sigma$ level (relative to statistical errors). However, extinction, as seen in the uppermost violin in Fig.~\ref{fig:violin_offax}, and initial lorentz factor, have much greater biases as they are largely informed by the optical/NIR light curve.
These inconsistencies again, prompt a review of methods to obtain more accurate predictions so that one may better interpret the afterglow. 

If we choose to subtract the best-fitting afterglow from the data to isolate the kilonova, the resulting inference not only still fails to correctly recover the input value of ejecta mass, but makes the estimation worse as can be seen by the red violins in Fig.~\ref{fig:violin_offax}. Instead, we may again turn to the use of a combined model. We show the joint fit in panel F of Fig.~\ref{fig:lightcurves}, and note that it provides a good visual fit to the true curves of both afterglow and kilonova components.
The full simultaneous inference results show that every parameter is not only recovered correctly, but the predictions have also improved in accuracy. Relative to statistical errors, all afterglow properties are inferred correctly to within $2\sigma$ and all kilonova properties are within $3\sigma$. We include the complete corner plot for all inferred parameters in Fig.~\ref{fig:sig_off_joint_corner} in the appendix. These results provide additional evidence that jointly fitting a contaminated signal is the most accurate and reliable method to use for disentangling and interpreting these transients.

\subsection{Scenario 3 - Data without X-ray and radio bands}
\label{sec: scenario 3}
\begin{figure}
    \centering
    \includegraphics[width=\linewidth]{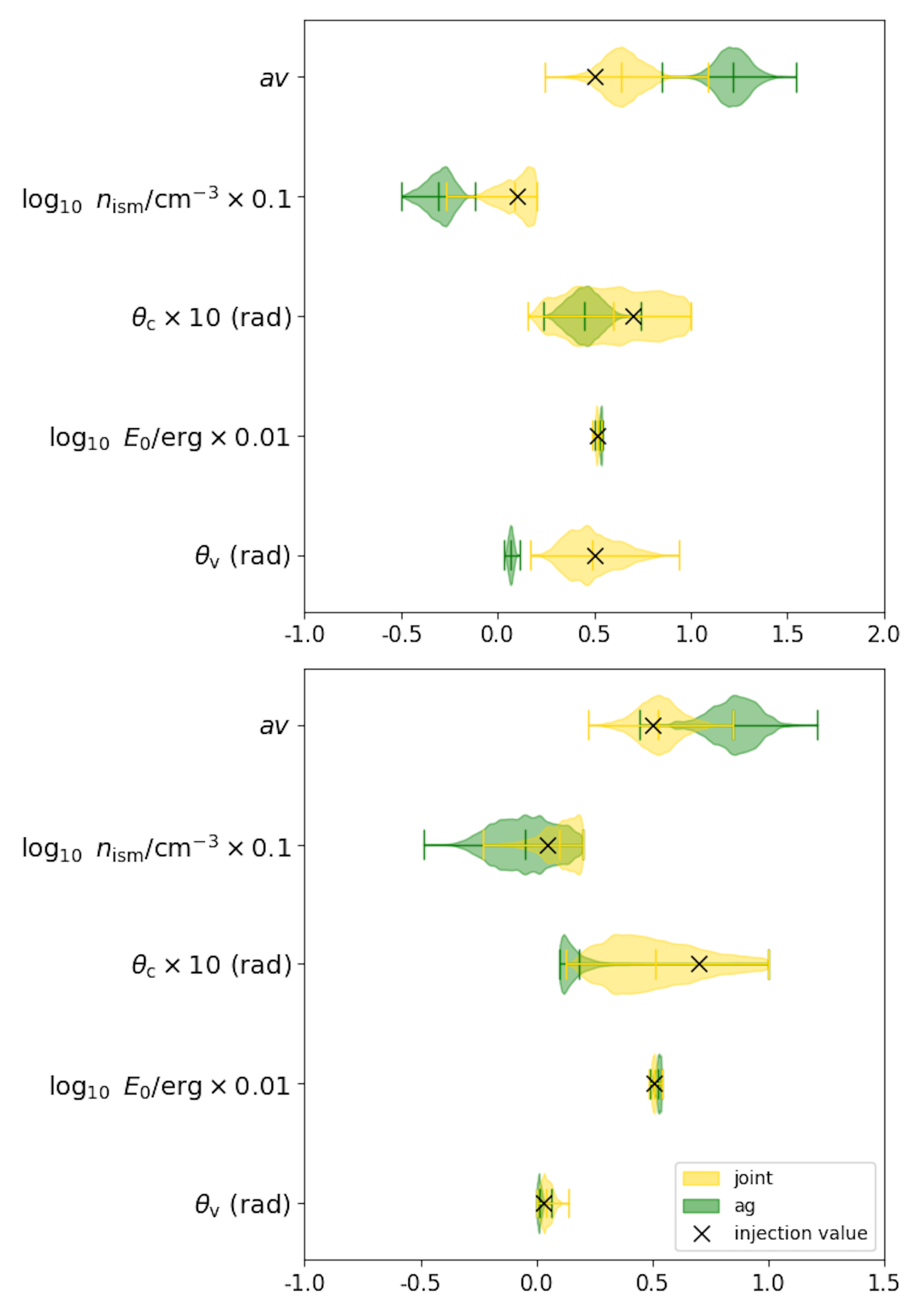}
    \caption{Equivalent afterglow violins to Figs.~\ref{fig:violin_onax} and \ref{fig:violin_offax} for the on-axis (lower panel) and off-axis (upper panel) recoveries missing X-ray and radio data. }
    \label{fig:violin_missing}
\end{figure}

We have surmised that the individual afterglow recoveries thus far benefit from the presence of X-ray and radio observations that remain free of kilonova contamination. Here, we explicitly test this by starting with the same data sets used in Sec.~\ref{sec: scenario 1} and Sec.~\ref{sec: scenario 2}. We first remove the X-ray and radio data points, and then reattempt to recover the afterglow component with our different fitting strategies. The result of this individual afterglow fit for the on-axis system (green violins in the lower panel of Fig.~\ref{fig:violin_missing}), without the extra bandpasses for calibration, correctly predicts most parameter values to within 3$\sigma$ relative to statistical errors. However, for many parameters the magnitude of the statistical uncertainties and therefore the width of the posterior distributions have increased relative to the individual inference of Sec.~\ref{sec: scenario 1}. This suggests that without X-ray and radio observations to `guide' the inference, the precision of some posteriors are decreased. 

\begin{figure*}
  \centering
  \includegraphics[width=\textwidth]{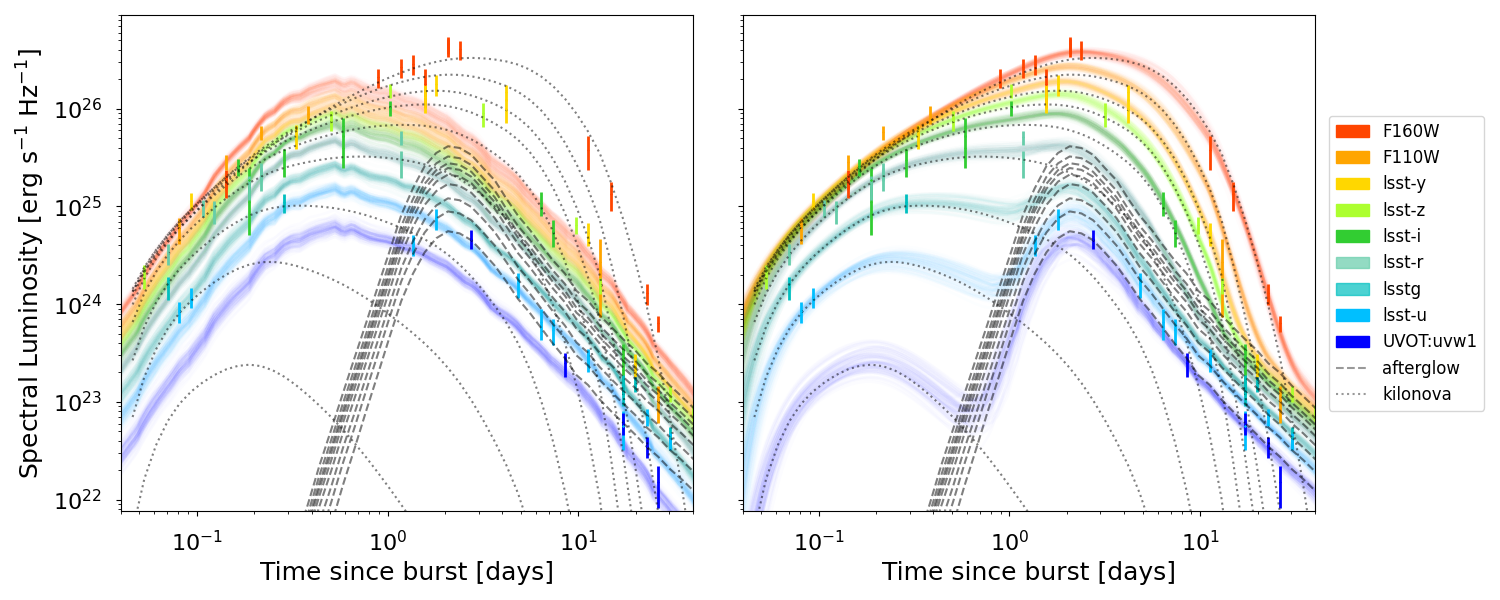}
  \caption{Predicted light curves for the off-axis contamination scenario without radio and X-ray data for an individual afterglow fit (left) and a joint fit (right). } \label{fig:optical_lightcurves}
\end{figure*}

The value of the uncontaminated X-ray and radio bandpasses is further emphasised when we consider the individual afterglow fit for the off-axis system. The afterglow only light curve fit, as presented in the left panel of Fig.~\ref{fig:optical_lightcurves}, is far from the ground truth and only $p$, $\epsilon_e$ and $\epsilon_B$ are recovered correctly to within 3$\sigma$. While the green violins showing the posteriors of select afterglow parameters in the upper panel of  Fig.~\ref{fig:violin_missing} seem to also show a successful recovery of $E_0$, the statistical errors are so small that the median value of the distribution is inconsistent with the input.
 
As the true afterglow light curve shape is difficult to determine from the data, and the X-ray and radio observations are not available to provide such information, the initially kilonova dominated data points appear to be fit as an extension of the afterglow signal itself. We saw in Fig.~\ref{fig:lightcurves} (panel E) from Sec.~\ref{sec: scenario 2} that the equivalent kilonova only recovery also fits these same data points and correctly infers the transient parameters. Since the individual fits contradict each other, if confronted with real data lacking the radio and X-ray observations, it could be difficult to determine which of the individual transient recoveries are correct. This means in the case of an off-axis system, which may have contamination and is more likely missing crucial X-ray and radio observations, individual fits or those performed after subtraction are not reliable.

Given the stark failure of individual fitting methods to the off-axis system, we apply the joint fitting strategy to both the on and off-axis scenarios, and compare our results. In both cases, we see improvement and find that all inferred parameters are consistent with the input values, as demonstrated by the yellow violins in Fig.~\ref{fig:violin_missing}. The use of a combined model allows a correct separation of the data points into their respective transient components, which resolves the problem observed in the individual fit, as shown in Fig.~\ref{fig:optical_lightcurves}. However, we note that while accuracy has improved, many of the posterior distributions have broadened (as expected) compared to the joint fits from Secs.~\ref{sec: scenario 1} and \ref{sec: scenario 2}, which included X-ray and radio observations. An example of this can be seen in Fig.~\ref{fig:sig_off_joint_optical_corner} of the appendix, where we show the full corner plot from the joint inference of the off-axis GRB system.

Another benefit of obtaining X-ray and radio observations is that it reduces the computational expense of Bayesian inference. Without the uncontaminated afterglow data to provide a `guide` when fitting, the sampling process is significantly longer as the efficiency of finding suitable draws from the prior is lower than it would be with the X-ray and radio data included. With MCMC methods where the number of posterior draws can be chosen, we found that often longer chains were required to produce satisfactory convergence. In this section, we therefore conclude that obtaining X-ray and radio data holds great value in terms of obtaining higher precision inference and reducing computational expenses. However, where it is not possible to obtain these observations, particularly in off-axis scenarios where prompt emission accompanied by X-ray and radio data is less likely to be detected, we emphasise that a joint fit is most likely the only way to correctly interpret the properties of the kilonova and afterglow. Additionally, we find that joint fits are less susceptible to problems associated with multi-modal distributions when using MCMC sampling. Bayesian inference with the combined model therefore proves to be the most suitable choice across all scenarios. 

\subsection{Scenario 4 - On-axis afterglow with little kilonova contamination}
\label{sec: scenario 4}

Thus far, we have shown that significant contamination between an afterglow and a kilonova can cause notable biases in parameter inference if the most common individual fitting strategies continue to be used. Here, we consider the scenario where contamination is much less significant. In this scenario, data is simulated for an on-axis GRB system with our key afterglow parameters changed to  $\log_{\rm 10}(E_0)= 52.0$, $\log_{\rm 10}(n_0)=1.5$, $\theta_v=0.03$~rad and $\theta_c=0.08$~rad, creating a particularly bright transient. The remainder of the injection values for the afterglow and kilonova remain the same as in the previous scenarios. This results in an afterglow dominated situation with very little contamination from the kilonova, which is observable as a small bump on the light curve at frequencies less than that of the lsst-z band.

\begin{figure}
    \centering
    \includegraphics[width=\linewidth]{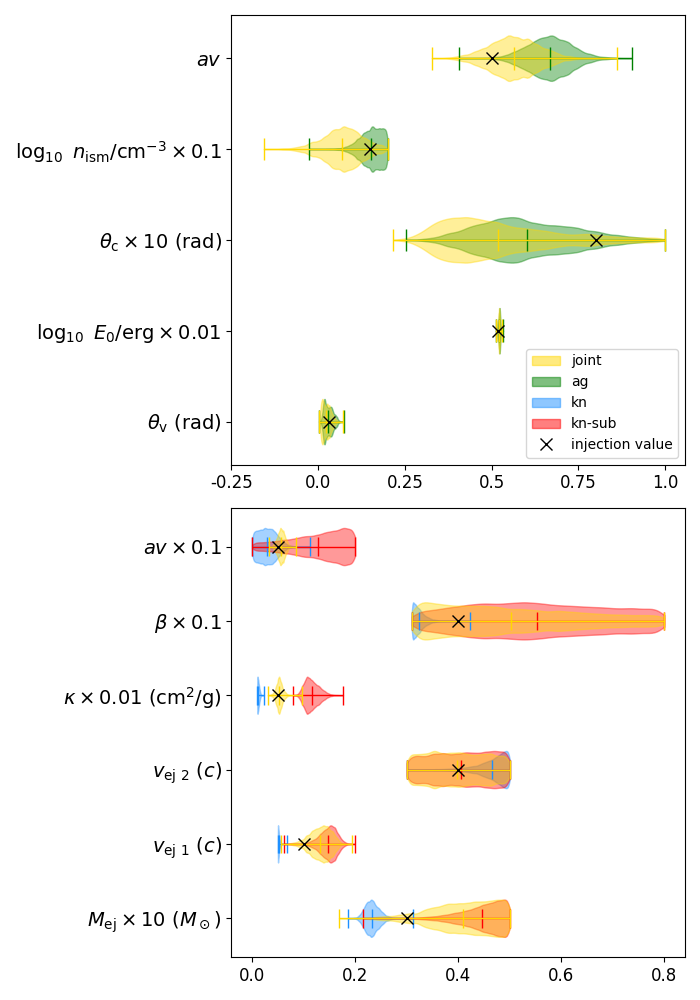}
    \caption{Similar to Fig.~\ref{fig:violin_onax} but for an on-axis GRB system where there is little contamination to the afterglow from the kilonova.}
    \label{fig:violin_agdom}
\end{figure}

\begin{figure}
    \centering
    \includegraphics[width=\linewidth]{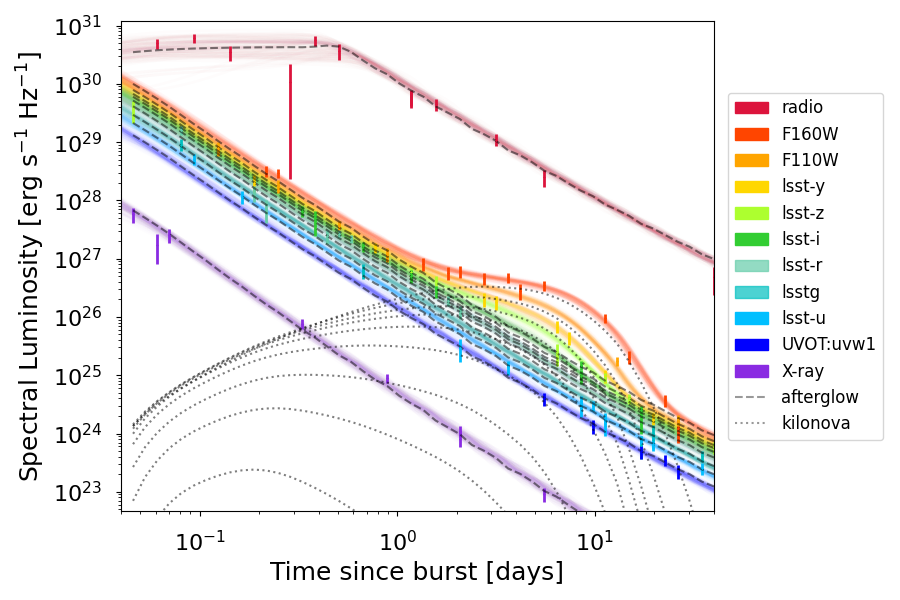}
    \caption{Predicted light curve for the joint fit of the afterglow dominated on-axis GRB scenario.}
    \label{fig:agdom_joint}
\end{figure}

Beginning, as usual, with the individual fits, we show in Fig.~\ref{fig:violin_agdom} the violin plots for the inference of key afterglow parameters in green and kilonova parameters in blue. These plots show that as expected, there are strong biases in the kilonova individual fits while estimates from the afterglow only fit are consistent with injection values. Given the small errors in the afterglow fit, one could again be motivated to consider the subtraction method. However, as shown in Fig.~\ref{fig:violin_agdom}, this methodology fails to correctly recover the opacity. By employing Bayesian inference with the joint model, we see from the yellow violins that the inferred properties of the kilonova become consistent with the injection values and importantly, remain consistent for the afterglow. The best-fitting posterior light curve for the joint fit is shown in Fig.~\ref{fig:agdom_joint} which highlights how subtle the contamination is compared to panel C in Fig.~\ref{fig:lightcurves}. Our results emphasise that where contamination is truly present, even if not obvious, a combined model is required to allow an accurate interpretation of the transients.

\subsection{Quantifying the robustness of fits assuming true contamination}
\label{sec: biases}

Situational differences between scenarios in this paper, like viewing axis and data availability, are more easily comparable because we only minimally vary the injection parameters of the afterglow. 
In this subsection we quantify the robustness of joint fitting across a wider variety of potential scenarios than have been investigated so far. Importantly, we also account for different data realisations of the same scenario that could impact the posterior distributions.
To accomplish this, we have selected six new sets of parameters that probe a larger range of the kilonova and afterglow prior space described in Tabs. \ref{tab:kn_priors} and \ref{tab:ag_priors}, and simulated 10 data realisations for each set of parameters. This totals 60 different synthetic observations, each of which we perform all fitting strategies on: afterglow only, kilonova only, kilonova only after afterglow subtraction, and joint.

\begin{figure*}
    \centering
    \includegraphics[width=\textwidth]{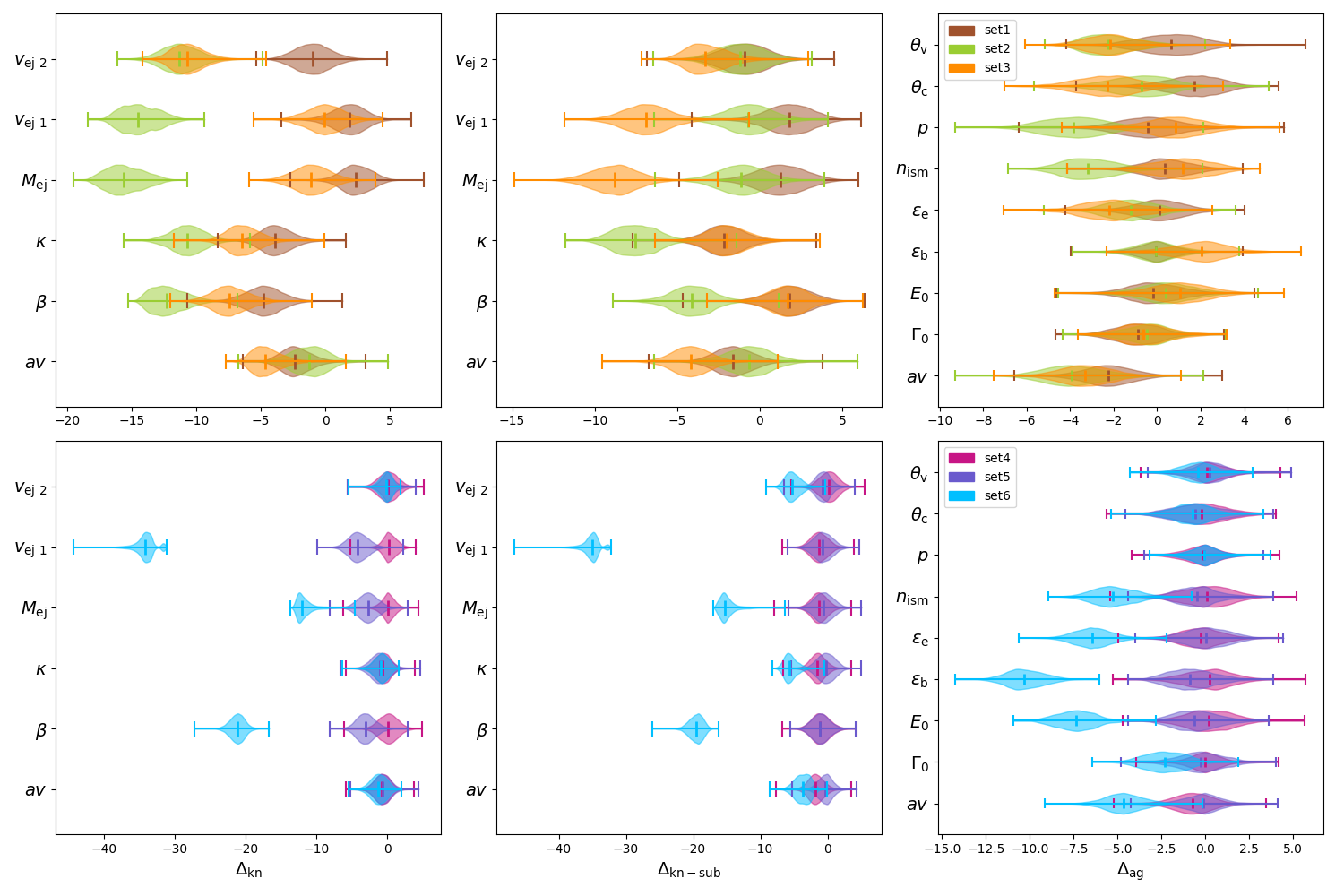}
    \caption{Sets 1-3 in the upper panels correspond to on-axis observation cases and sets 4-6 in the lower panels correspond to off-axis observation cases. Each subplot contains violins showing the average distribution of error $\Delta$ across 10 different data realisations per set, for each parameter in units of average standard deviation relative to the joint fit method. From left to right, columns show results relative to the joint fit for the kilonova only fit, kilonova only fit after afterglow subtraction, and the afterglow only fit.}
    \label{fig:bias_violins}
\end{figure*}

The parameters for each data set are still selected with the requirement that the afterglow and kilonova contaminate, and we intentionally use three scenarios where the afterglow is viewed on-axis and three where the afterglow is viewed off-axis. To compare the biases between each fitting method, we perform the following analysis on each of the six sets of data. For every recovery method, we first average over the posterior distributions from the 10 different data realisations which allows us to calculate the standard deviation $\sigma_{avg}$ of the average posterior distributions. We then calculate the absolute difference $\delta$ between the average of the posteriors for each parameter and the true injection values. The absolute difference is expressed in units of $\sigma_{avg}$. When these analysis steps are completed for each recovery method, we calculate the difference $\Delta$ between $\delta$  for the joint method and all other methods such that $\Delta_{method} = \delta_{joint}-\delta_{method}$. Using this metric, a more negative (positive) $\Delta$ indicates the parameter values inferred using the specified method of recovery are $\Delta  \times \sigma_{avg}$ worse (better) than the joint recovery method. The results of this analysis are presented in Fig.~\ref{fig:bias_violins} where the subscripts `kn',  `kn-sub' and `ag' refer to the kilonova only, kilonova only after afterglow subtraction, and afterglow only fitting methods, respectively. Different colours indicate different sets of data where sets 1-3 are contamination scenarios of on-axis afterglows and sets 4-6 are contamination scenarios of off-axis afterglows.

A violin perfectly centered about $\Delta=0$ would indicate that the joint fitting method infers parameter values to the same degree of accuracy as the individual fitting methods. Compared to the kilonova fits, the afterglow parameter violins across both on and off-axis cases are generally centered closer to 0, suggesting that they can give comparable results to the joint fit, as we determined was a consequence of including X-ray and radio data. However, there are some notable outliers, especially in set 6, which indicate that the afterglow only fit can be up to 10 standard deviations worse than the joint fit at predicting the values of certain parameters, even with the inclusion of data from these crucial bandpasses.
We also note that set 6 shows significant outliers in kilonova parameter space compared to the other off-axis contamination scenarios, with statistical inaccuracies from the kilonova only fit fairing more than 30$\sigma$ worse than the joint fit counterpart at the extreme. 

As previously discussed, the kilonova only recoveries are expected to perform better in off-axis contamination scenarios than on-axis ones because the kilonova is dominant at early epochs. While Fig.~\ref{fig:bias_violins} shows that this is generally true, with the $\Delta$ values from sets 1-3 tending to be more negative than sets 4 and 5 on average, the severe outliers in set 6 demonstrate that an off-axis afterglow is an insufficient condition to make either the individual or afterglow subtracted methods for kilonova inference trustworthy options. Furthermore, we note that while the afterglow subtracted kilonova fits generally show more positive $\Delta$ values compared to kilonova fits without prior data subtraction, we see an exception to this in set 3 where the $\Delta$ values for $v_{ej-1}$ and $M_{ej}$ become more negative. This shows that using the subtraction technique does not guarantee an improvement in accuracy relative to purely individual fitting techniques.

\subsection{Scenario 5 - Model misspecification with weakly-informative data and no kilonova}
\label{sec: scenario 5}

In this final scenario, we test the limits of our methodology and simulate an afterglow observation with no contamination i.e., there is no kilonova. This scenario is particularly relevant to claims of kilonova detections when understanding the distinction between long and short GRBs. The presence of a kilonova in the previous scenarios was well constrained given high time sampling in each bandpass, so we now reduce the number of data points to 20 and limit our observations to radio, X-ray, and LSST irgu bands. We also raise the level of noise in the observations to mimic a lower signal-to-noise observation. We simulate the afterglow with $\log_{\rm 10}(E_0)= 50.5$, $\log_{\rm 10}(n_0)=-1.0$, $\theta_v=0.04$~rad and $\theta_c=0.08$~rad. In our analysis, we still attempt all types of fit as previously to see if a kilonova can be predicted when not actually present.

\begin{figure}
    \centering
    \includegraphics[width=\linewidth]{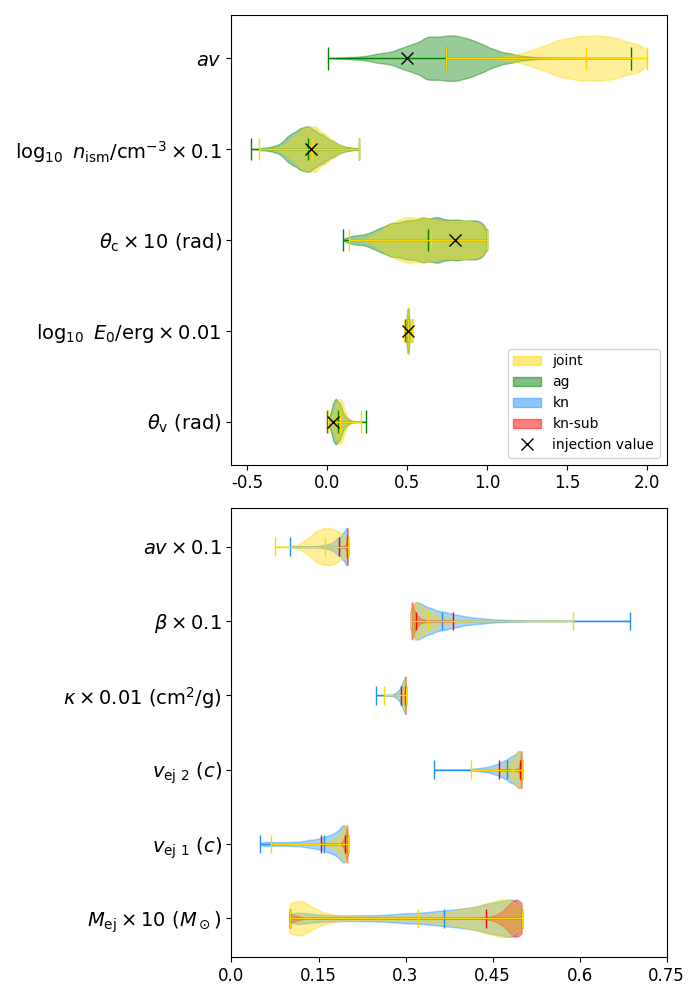}
    \caption{Similar to Fig.~\ref{fig:violin_onax} but for an afterglow with no true kilonova contamination, and therefore no injection values in the lower panel.}
    \label{fig:violin_nokn}
\end{figure}

\begin{figure*}
    \centering
    \includegraphics[width=\linewidth]{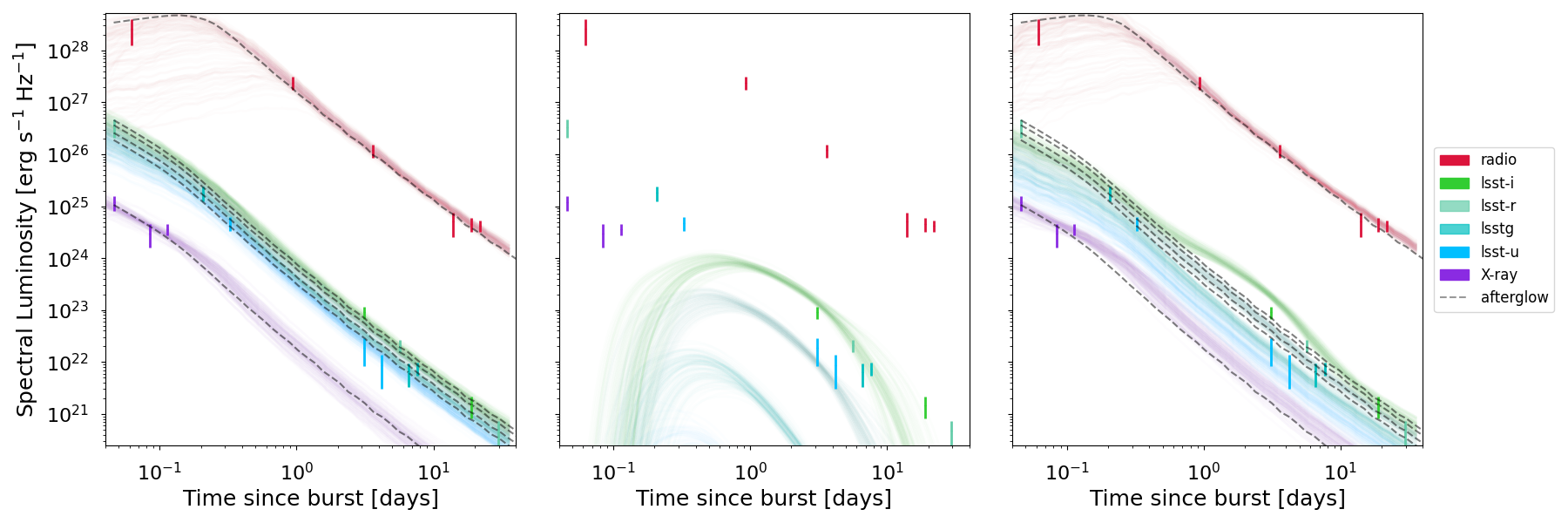}
    \caption{Posterior light curves for the afterglow only fit (left), kilonova only fit (middle), and joint fit (right), for the on-axis afterglow scenario with no true kilonova contamination and uninformative data.}
    \label{fig:no_kn_lightcurves}
\end{figure*}

Fig.~\ref{fig:violin_nokn} shows that since there is no longer contamination from a kilonova, the individual afterglow fit provides accurate results, inferring parameter values within 1$\sigma$ of the injection values. The posterior light curve for this fit is shown in the left panel of Fig.~\ref{fig:no_kn_lightcurves} which matches well to both the data and the ground truth curves. In this low quality data scenario, we find that the parameters relating solely to the afterglow are also recovered using the joint fit as reflected by the yellow violins in Fig.~\ref{fig:violin_nokn}. However, extinction, being a shared parameter between the afterglow and kilonova in the joint fit, is biased by the attempt to recover the latter non-existent transient and shows no posterior overlap with the true input. More generally, and especially when the data is informative (more bandpasses and higher time sampling), the independent afterglow fit is the only method that can reliably recover the ground truth. This is again a product of model misspecification, i.e., a bias induced in Bayesian inference when models used in fitting  are not representative of the true data generation process.

When we remove the best-fitting afterglow from the data, there are very few points remaining to perform the subtracted kilonova fit on. The resulting inference, shown by the red violins in Fig.~\ref{fig:violin_nokn}, produces posteriors for all parameters that are heavily skewed to the edges of our prior space and have small or no associated statistical error, as the posteriors are strongly railing against the prior. Simply, in such a scenario there are not enough remaining points to perform a good fit and, as one might expect, the posterior light curves do not match the data well.

We find that both the individual kilonova fit and the joint fit predict a kilonova, despite that it is not present in the data, but whether the fit is convincing is highly dependent on the quality of the data. In this scenario, the data is particularly uninformative in the redder filters and between approximately 0.4 – 3 days, which aligns with the optimal kilonova observation regime that we highlighted in Sec.~\ref{sec: Regime of Contamination}. The resulting light curve fits shown in Fig.~\ref{fig:no_kn_lightcurves} from both the individual (middle panel) and joint (right panel) fits seem sensible for this scenario, with respect to the data. However, with higher quality data in the aforementioned regime, fits can again lead to posteriors that strongly rail against the prior and posterior light curves that would raise suspicion – indicators of a misspecified model. We note that in cases like the example we have shown, where data is uninformative, the prediction of a kilonova using the joint fit may be a concern for interpreting results. In these situations, posterior predictive checks could help to determine the misspecified model. We do not detail this further here, but instead refer the reader to e.g. \cite{Gelman1996} and references therin.

\section{Conclusions}
\label{sec: conclusions}
Detailed observations are beginning to readily reveal contamination of a GRB afterglow with additional components. In this work, we have explored several strategies for fitting such observations and inferring the properties of the individual afterglow and additional component. In particular, we have focused on the scenario of a kilonova and an afterglow motivated by recent observations of long GRB afterglows with infrared excesses consistent with a kilonova~\citep{GRB211211A, GRB230307A, Yang_GRB230307A}. 

We first determine the region of parameter space (time, wavelength and afterglow and kilonova properties) where contamination is more likely.  In general, kilonova emission is brighter and lasts longer in the lower frequency NIR bands, up to $\unit[30]{days}$ post GRB, in comparison with higher frequency UV bands where the onset of decay can begin as soon as $\unit[1]{day}$ post GRB. In an on-axis contamination scenario, the kilonova will be observed as a bump in the lightcurve, the significance of which is sensitive to the spectral luminosity of the afterglow component. The optimal observation window in this case is approximately $\unit[2-15]{days}$.  In an off-axis scenario, the kilonova will always initially dominate the signal, which we have shown is generally favourable for kilonova parameter estimation. The takeaway, regardless of viewing angle, is to observe as early as possible and make use of NIR bands if aiming to detect a kilonova, which is consistent with current observational strategy. 

We then explored different strategies to fit the data and estimate the parameters of the kilonova and afterglow using synthetic, simulated data where we know the true input. We apply the different strategies to different contamination scenarios, and account for changes in posterior distributions due to different random noise and data realizations within the same scenario. In many cases where the afterglow and kilonova are fit individually, we find that a predicted light curve may look reasonable to the eye while the estimated parameters are inconsistent with the true input, highlighting the dangers of this strategy for correct interpretations. In our on-axis observation scenarios, the inferred kilonova cannot be recovered using individual model fitting, and we find this holds true across different parameter variations and data realisations. For off-axis observations, the kilonova is generally fit better due to its dominance in optical bandpasses at early times, but exploring different parameter combinations shows that this does not rule out the possibility of encountering severe biases with an individual model inference approach. The more commonly used strategy of subtracting a best-fitting afterglow and then fitting a kilonova fares better, but again the kilonova parameters can be biased and in some cases can even get worse. We note that this also does not produce a fully marginalised posterior, but instead a posterior on the kilonova parameters conditioned on the afterglow model and best-fitting parameters, which is not ideal. We find that the best strategy across all observation angles is to jointly fit both the afterglow and kilonova, and have showcased a Bayesian framework (publicly available via {\sc Redback}) to perform such joint analyses. The resulting inferred values using the joint method can be at worst comparable to individual methods and at best better than 30$\sigma$ more accurate.

When the afterglow is missing observations in the radio and X-ray bands, it can be difficult to distinguish between data points that are strictly from afterglow contributions and those that are strictly from kilonova contributions. This is especially true for off-axis systems where data points from the initial kilonova-dominated signal can be fit as if they were an extension of the afterglow. Individual parameter inference is unreliable in these scenarios and is difficult to improve via a subtraction technique due to contradicting fits between the afterglow and kilonova. The only solution here is to again use a joint-fitting approach. However, while X-ray and radio data are not necessary to ensure a successful joint fit, these observations hold the power to decrease computational time by improving sampling efficiency and significantly improve the precision of predictions and, therefore, should remain a priority when observing.

In the limiting case where an afterglow is observed but there is no true kilonova contamination, we find that the independent afterglow fit is the only method that will reliably recover the correct parameters due to biases induced from a model misspecification, highlighting the need to perform posterior predictive checks that can test for biases \citep[e.g.,][]{Gelman1996}. The afterglow properties from a joint fit are more likely to have an accurate (biased) recovery when the data is less (more) informative within the window where kilonova observation is favourable. Likewise, when the data is less (more) informative in this regime, the joint and individual kilonova fits can predict a transient that appears (un)reasonable. Kilonova fits that do not appear sensible, including those resulting from afterglow subtraction when very few data points remain, often have posteriors that strongly rail against the priors and may serve as an indicator of a misspecified model. We note that when data is of low quality, caution should be taken when interpreting the results of a joint fit that predicts a kilonova, even one that appears to have reasonable parameters.

In this paper, we have demonstrated that where there is contamination between a kilonova and an afterglow, using a joint model for simultaneous transient inference is the ideal methodology for accurate and robust parameter estimation. Joint fitting also proves to be the most reliable method for inference, succeeding in situations where individual fits or subtraction techniques fail. While the combined model has been successful on our synthetic data sets, we recognise that there is significant room for development to make the combined model more realistic. In particular, our combined model implicitly assumes that the afterglow and kilonova emission is uncoupled, contrary to numerical simulations that show that kilonova ejecta properties change due to the traversal of a jet~\citep{Klion2021, Nativi2021}. Furthermore, our analytic models are limited by the assumption of a constant, grey opacity, which implicitly builds a shape to the overall kilonova light curve that numerical simulations of kilonova light curve do not always obey. An alternative approach to the semi-analytical models we use would be to use machine-learning based surrogates to model the numerical simulations of kilonovae~\citep[e.g.,][]{kilonovanet}. However, these simulations are performed under the assumptions of local thermal equilibrium (LTE), which is invalid for most kilonova parameters past $\approx \unit[10]{d}$~\citep{nlte}, and can have a significant impact on the light curves past this epoch. Few studies have performed non-LTE kilonova modelling due to dramatically higher computational cost and nuclear data required to compute such light curves. As this work uses only simulated data, non-LTE effects are not considered, and the simplified semi-analytical formulae for kilonovae are considered the `true' input for our analysis, even past $\unit[10]{d}$, which could affect our conclusions. Further detailed modelling may also account for an afterglow jet with angular structure. Both such changes may help with the accuracy of parameter estimation, but it is unlikely that they will significantly change our conclusions about the ideal choice of methodology. 

An important caveat to our exploration of the idealised approach to fitting is the consideration that the models themselves are also inherently uncertain. While the choice of fitting strategy, e.g. subtraction, may bias our results, the bias is (at least in most of our simulated scenarios) smaller than the systematic uncertainty in the kilonova models themselves due to uncertain underlying physics~\citep{sarinrosswog24, Brethauer2024}. Therefore, we expect broad (order of magnitude estimates) of kilonova properties from subtraction-based fitting strategies~\cite[e.g.,][]{GRB211211A, jillian24} are likely robust. However, the bias is likely more of a concern for the relatively more certain afterglow parameters. We leave a detailed exploration with real observations for future work. We note that regardless of the size of bias a fit based on individual component fitting or subtraction creates, it is not the ideal quantity one desires when estimating parameters through Bayesian inference; a marginalised posterior distribution, dependent only on the observed data. Instead, these alternative approaches implicitly condition the posterior on the choice and parameters of a best-fitting afterglow model or some other underlying assumption. This is not ideal when what we truly desire is a posterior marginalising over every other degree of freedom. 

\section*{Acknowledgements}
W. Wallace would like to thank Nordita for hosting her visit for the duration of this project, and acknowledges support for the mobility from a Turing Scheme grant awarded through the University of Bath, and support from the National Aeronautics and Space Administration (NASA) under award number 80GSFC24M0006. Any opinions, findings, and conclusions or recommendations expressed in this material are those of the authors and do not necessarily reflect the views of NASA. N. Sarin acknowledges support from the Knut and Alice Wallenberg Foundation 
through the "Gravity Meets Light" project and by the research environment grant
``Gravitational Radiation and Electromagnetic Astrophysical
Transients'' (GREAT) funded by the Swedish Research Council (VR) 
under Dnr 2016-06012 and the support through the Nordita Fellowship. Nordita is supported in part by NordForsk.

\section*{Data Availability}
Analysis and fitting is carried out in {\sc python} using the open source software package {\sc redback}~\citep{redback_paper}, with the data generated via the simulation interface of this package. The models used/developed for this project are implemented in {\sc redback} and the neural network is implemented in {\sc redback\_surrogates}. Both software packages are available on {\sc{pypi}}, with documentation at \url{https://redback.readthedocs.io/en/latest/} and \url{https://redback-surrogates.readthedocs.io/en/latest/} respectively.



\bibliographystyle{mnras}
\bibliography{ref} 



\appendix
\renewcommand{\thefigure}{A\arabic{figure}}
\setcounter{figure}{0}
\section*{Appendix A} \label{appendix}

Here we include the full posteriors for different analyses performed in this work.
\begin{figure*}
    \centering
    \includegraphics[width=\textwidth]{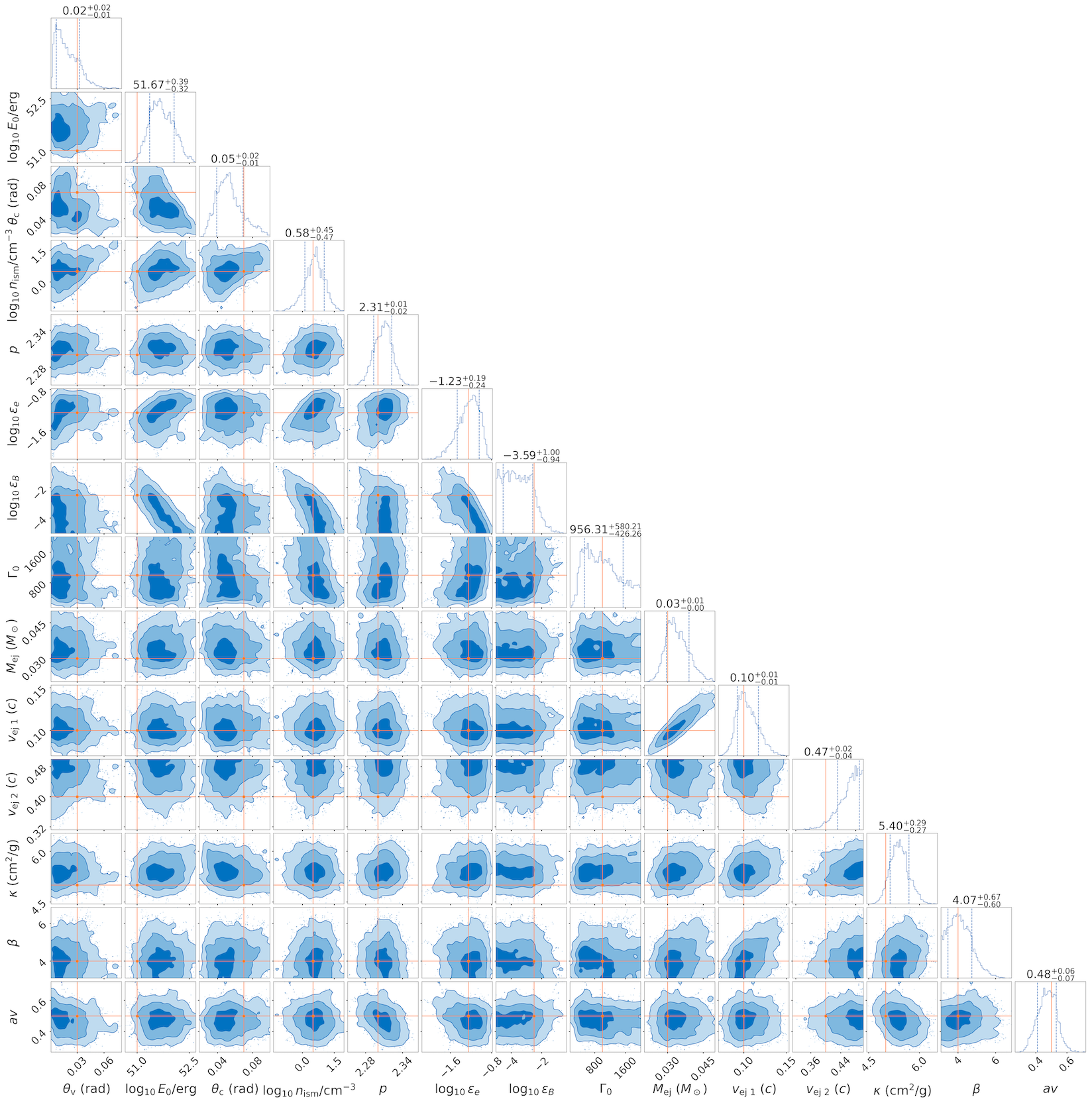}
    \caption{Corner plot for a joint fit to the on-axis contamination scenario. Along the diagonal are the 1D marginalised posterior distributions for each parameter, and the blue contours show the credible intervals up to 3$\sigma$ for the 2D posteriors. The true input values are denoted by the orange lines.}
    \label{fig:sig_on_joint_corner}
\end{figure*}

\begin{figure*}
    \centering
    \includegraphics[width=\textwidth]{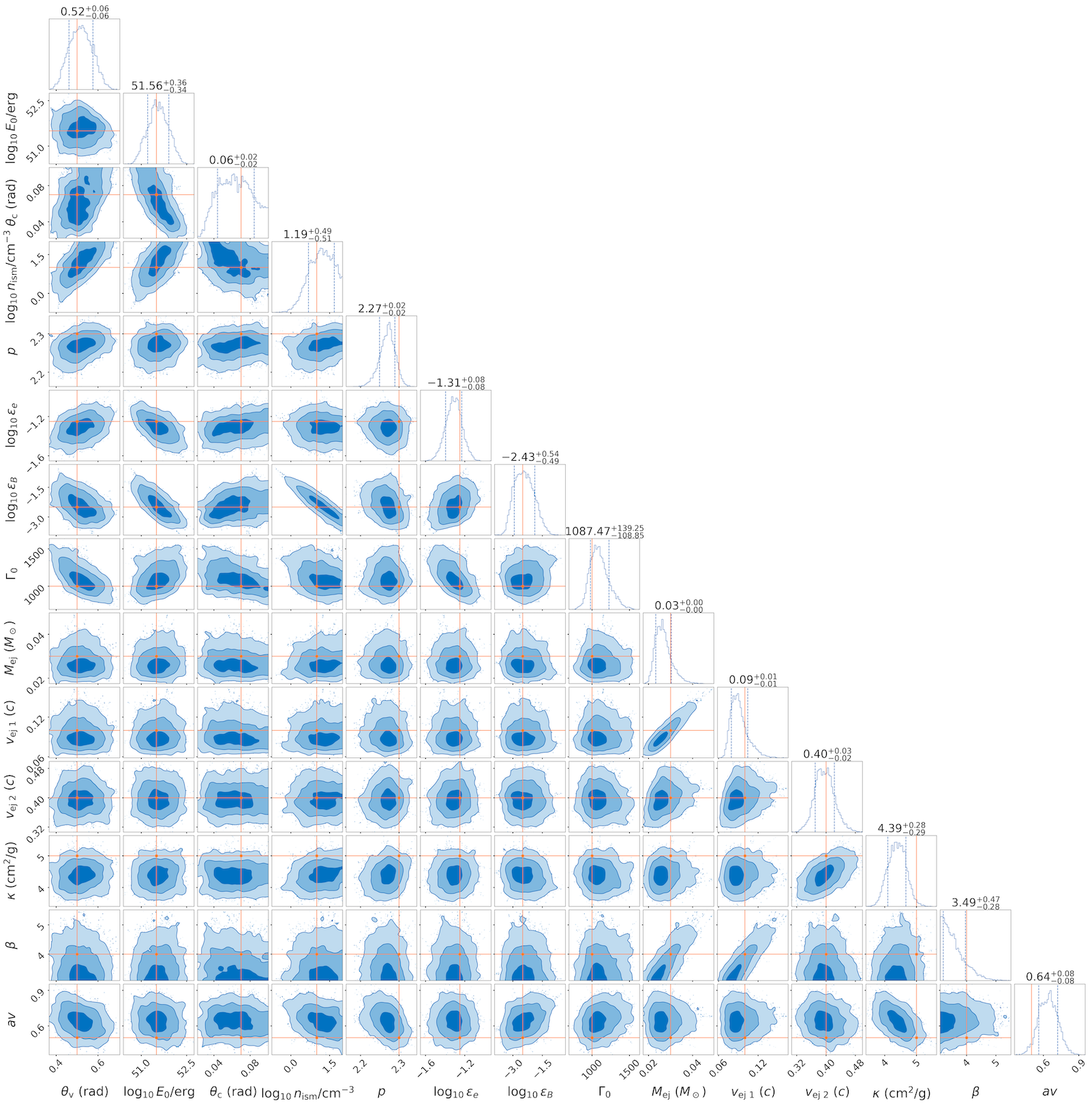}
    \caption{Similar to~\ref{fig:sig_off_joint_corner} but for the off-axis contamination scenario.}
    \label{fig:sig_off_joint_corner}
\end{figure*}

\begin{figure*}
    \centering
    \includegraphics[width=\textwidth]{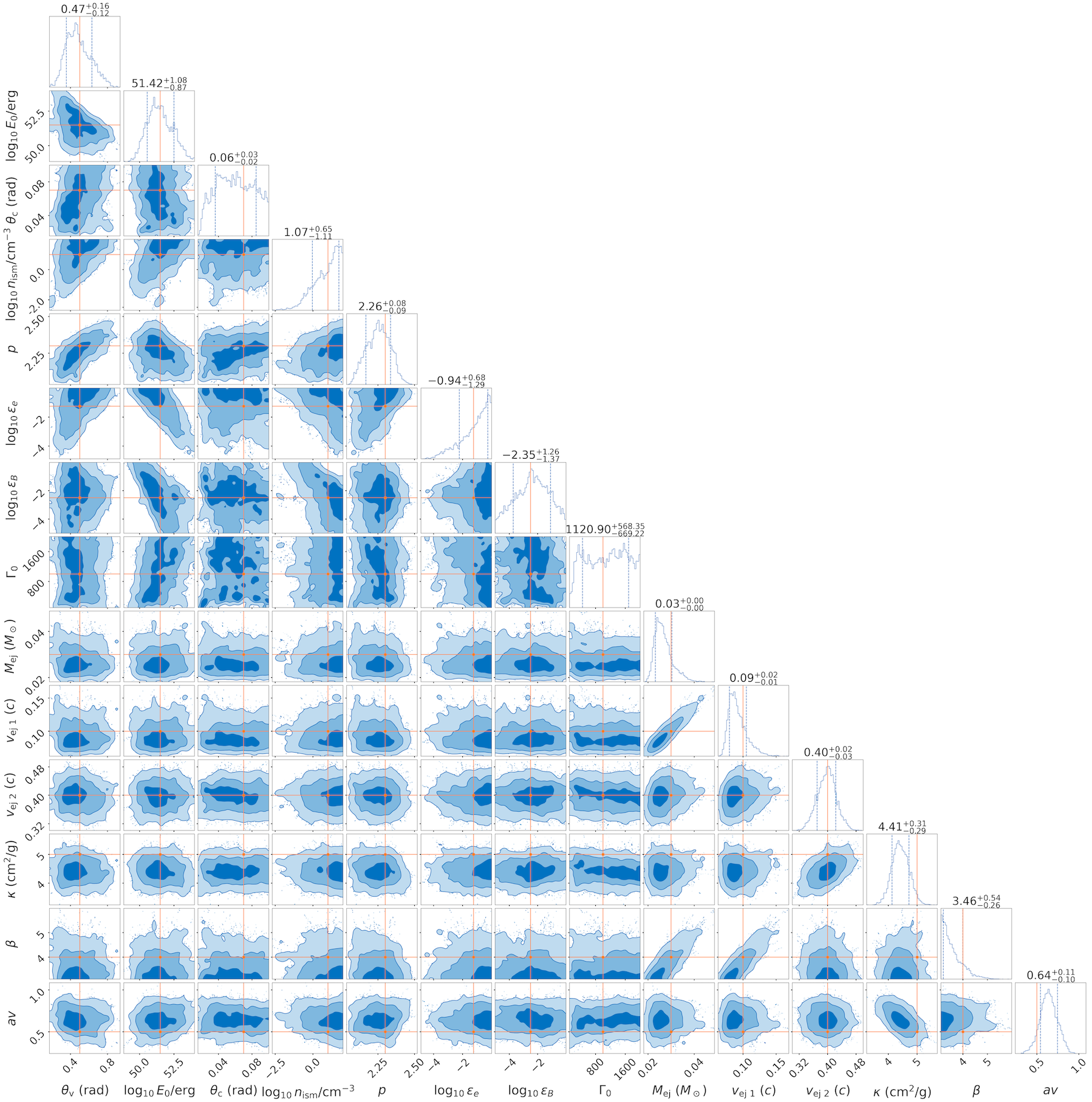}
    \caption{Similar to~\ref{fig:sig_on_joint_corner} but for the off-axis contamination scenario without X-ray and radio data.}
    \label{fig:sig_off_joint_optical_corner}
\end{figure*}

\bsp	
\label{lastpage}
\end{document}